# Magnetic structure of polar magnet GaV$_4$Se$_8$ with Néel-type skyrmion lattice probed by $^{51}$V NMR

Hikaru Takeda[1], Misaki Ishikawa[1], Masashi Takigawa[1], Minoru Yamashita[1], Yuri Fujima[2], and Taka-hisa Arima[2]

[1] *Institute for Solid State Physics, University of Tokyo, 5-1-5 Kashiwanoha, Kashiwa, Chiba 277-8581, Japan*

[2] *Department of Advanced Materials Science, University of Tokyo, 5-1-5 Kashiwanoha, Kashiwa, Chiba 277-8561, Japan*

**Abstract**

We report the magnetization and the $^{51}$V NMR measurements in the polar magnet GaV$_4$Se$_8$ in which a magnetic skyrmion lattice appears in the structural domain with the polar axis parallel to the magnetic field. Although we successfully separate the $^{51}$V NMR signals in the domain from those in the other structural domains, only the high-frequency region of NMR spectrum is observed due to a significant increase of the spin-echo decay rate in the low-frequency region. In **B**$_{ext}$ ∥ [111], we find the NMR signals from the supermagnetized cycloidal state in the parallel domains as well as from the conical state in domains where the polar axis is tilted from the magnetic field. No NMR signal from the skyrmion lattice state is observed, suggesting a significant increase of the decay rate by additional low-energy excitations caused by dynamics of the skyrmions. In **B**$_{ext}$ ∥ [001], where all the structural domains are magnetically equivalent, multiple NMR peaks converge into one peak at the saturation magnetic field. This field dependence is explained by the closing of magnetic cones as approaching the forced-ferromagnetic state.

## I. Introduction

In noncentrosymmetric lattice systems, the antisymmetric exchange interaction called the Dzyaloshinskii-Moriya (DM) interaction [1,2] provides rich and complex magnetism. Combined with antiferromagnetic and/or ferromagnetic exchange interactions, the DM interaction stabilizes spatially modulated spin structures such as canted antiferromagnetic order, helical spin alignments, and magnetic vortices. A magnetic skyrmion is one of the prominent structures stabilized by the competition between symmetric spin-exchange interactions and DM interactions [3,4]. Due to the topological and chiral nature, the skyrmion exhibits intriguing phenomena such as topological Hall effects [5–7], easy manipulations by small electric current [8–10] or multiferroic behavior [11], which have attracted great attentions.

Several compounds have been reported to host a periodic lattice of the skyrmion. For example, chiral magnetic crystals such as MnSi [12], FeGe [13], Fe$_x$Co$_{1-x}$Si [14] and Cu$_2$OSeO$_3$ [15] host a Bloch-type skyrmion lattice (SkL) with whirlpool-like structure due to the chiral lattice symmetry. In these materials, ferromagnetic exchange interactions, which compete with the uniform antisymmetric exchange interactions, stabilize the skyrmion lattice in the magnetically ordered state. Another example is a magnet with a polar symmetry $C_{nv}$ such as GaV$_4$X$_8$ ($X$ = S, Se) [16–21] and VOSe$_2$O$_5$ [22], in which the symmetry permits Néel-type skyrmions [3]. In contrast to the chiral crystals, orientations of the magnetic propagation vectors are restricted to lie in the plane perpendicular to the polar axis, which affects the stability of the SkL state.

Recently, lacunar spinel compounds GaV$_4$X$_8$ ($X$ = S, Se), which undergo a structural phase transition from a noncentrosymmetric cubic structure at room temperature to a rhombohedral polar structure at low temperatures, have attracted special attention, since these compounds host a relatively stable SkL state against external magnetic fields and temperatures [16–21]. The stability of the SkL in GaV$_4$X$_8$ is ascribed to the lattice symmetry. The polar and achiral symmetry at low temperatures stabilizes the cycloidal magnetic order with a propagation vector $q$ confined in the plane perpendicular to the polar axis. Since the propagation vector $q$ is fixed in the plane, the out-of-plane field stabilizes a triple-$q$ state, namely the SkL. This is in contrast to the chiral magnets which host Bloch-type skyrmions. In the chiral magnets, the SkL phase competes with the longitudinal cone state, because the propagation vector $q$ can flop parallel to the external field. On the other hand, the SkL state in the polar magnet does not have such a competitive phase. Thus, GaV$_4$X$_8$ have a vast SkL phase in the $B_{\text{ext}}$-$T$ phase diagram. In particular, GaV$_4$Se$_8$ hosts a stable SkL phase, which persists down to the lowest temperature. GaV$_4$Se$_8$ exhibits a structural phase transition at $T_\text{S} = 41$ K, followed by a magnetic transition at $T_\text{N} = 18$ K. Below $T_\text{S}$, the cubic lattice is elongated along one of the ⟨111⟩ axes to change to a rhombohedral structure. This structural transition results in four kinds of crystallographic polar domains with the polarization direction along the four cubic-body diagonals (see $Z_\text{A}$–$Z_\text{D}$ in Fig. 1(a) and Table I) [19,23–25]. Application of an external magnetic field $\mathbf{B}_{\text{ext}}$ along the polar [111] axis below $T_\text{N}$ transforms the cycloidal state at $B_{\text{ext}} = 0$ to the Néel-type skyrmion lattice state in the polar domain with the polarization direction parallel to $B_{\text{ext}}$ (denoted as domain A), which is followed by the forced-ferromagnetic (FM) state above the saturation field of about 0.4 T [19,20]. In the other three types of domains (domains B–D), the cycloidal state is immediately modified to a conical state, which is followed by saturation above ~0.15 T. The appearance of the SkL in domain A is also observed as the emergence of the topological thermal Hall effect of magnons by the emergent magnetic field produced by the SkL [21]. It has been proposed that the easy-plane anisotropy of GaV$_4$Se$_8$ stabilizes the SkL phase down to the lowest temperature, as confirmed by the persistence of the thermal Hall effect in the SkL phase down to 0.2 K. The vast SkL-phase region in the $B_{\text{ext}}-T$ phase diagram of GaV$_4$Se$_8$ provides us an opportunity to investigate the local magnetic structure of the skyrmion.

In this study we investigate the magnetic field evolution of the magnetic structure of GaV$_4$Se$_8$ by $^{51}$V NMR measurements. Taking advantage of the site selectivity in the NMR measurements, we elucidate multiple magnetization processes separately in an external magnetic field from the analysis on the $^{51}$V NMR spectra. In $\mathbf{B}_{\text{ext}} \parallel [111]$, the NMR signals coming from domain A and the other domains (B–D) are observed separately. Although the NMR signals from the conical state in domains B–D persist up to the FM state, the NMR signal in domain A is observed only below 0.2 T in the magnetization process, which completely disappears in the demagnetization. From this magnetic hysteresis, as well as that observed in the topological thermal Hall effect [21], we conclude that only the NMR signal from the supermagnetized cycloidal state is observed and that the NMR signal from the SkL state is absent. We suggest that some additional low-energy excitations in the magnetic skyrmions may make the spin-echo decay time too short to detect the NMR signal. In $\mathbf{B}_{\text{ext}} \parallel [001]$, where the applied magnetic field is equivalent for all four structural domains, multiple NMR peaks are observed, which converges to a single peak above the critical field for the phase transition to the FM state. From the magnetic-field evolution of the spectral shape, we discuss two possible processes for a spin reorientation from the cycloidal to the FM state passing through the conical or skewed SkL states.

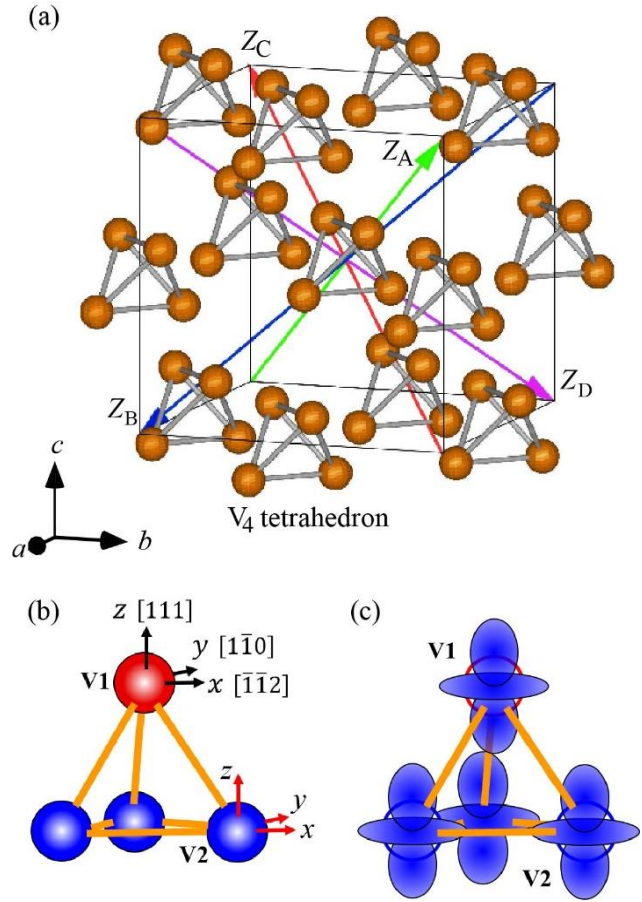

Fig. 1 (a) Schematic illustration for V$_4$ tetrahedra in the cubic unit cell of GaV$_4$Se$_8$ drawn by VESTA [26]. Four arrows denoted by $Z_A$–$Z_D$ are four possible polar axes in the rhombohedral phase below $T_S = 41$ K. Schematic figures of V$_4$ tetrahedra with (b) local coordinate axes and (c) the $a_1$ molecular orbital in which an unpaired electron is occupied in the low-temperature phase. The directions of the local axes are expressed for domain A in the cubic setting.

Table I. Orientations of the local axes in four structural domains of GaV$_4$Se$_8$.

| domain | X | Y | Z |
|---|---|---|---|
| A | $[\bar{1}\bar{1}2]$ | $[1\bar{1}0]$ | $[111]$ |
| B | $[\bar{1}1\bar{2}]$ | $[110]$ | $[1\bar{1}\bar{1}]$ |
| C | $[112]$ | $[\bar{1}10]$ | $[\bar{1}\bar{1}1]$ |
| D | $[1\bar{1}\bar{2}]$ | $[\bar{1}\bar{1}0]$ | $[\bar{1}1\bar{1}]$ |

## II. Experimental Procedure

Two single crystals, samples #1 and #2, were used for the magnetization and the [51]V NMR measurements. Both crystals were obtained by using a chemical vapor transport method as described in the previous paper [20]. Magnetization measurements were performed on sample #1 by using a superconducting quantum interference device (SQUID) magnetometer (Quantum Design, magnetic property measurement system (MPMS)). [51]V NMR measurements of sample #2 were performed by the pulse method. Sample #2 with a tetrahedral shape was placed in the NMR coil so that the radio-frequency (rf) field is applied along the $[1\bar{1}0]$ direction as shown in Fig. A1 (see Appendix A). In the magnetically ordered phase, the overall spectra were taken by recording the integrated intensity of the spin-echo signal at discrete frequencies, while a part of the spectra near the high-frequency edge was taken by summing the Fourier transform of the spin-echo signal obtained at equally spaced frequencies (Fig. 5). Since the spin-echo decay time $T_2$ for the nuclear spin at [51]V is very short, we set the duration of the first and second rf pulses at $0.8$ μs and a delay time $\tau$ which separates these two pulses as short as possible. The pulse-voltages are calibrated to maximize the spin-echo intensity as shown in Appendix A. The spin-echo decay rate $1/T_2$ was evaluated by fitting the echo intensity $I(2\tau)$ as a function of $2\tau$ to a single exponential function with the time constant $T_2$.

The [51]V NMR frequency in GaV$_4$Se$_8$ is governed by the nuclear spin Hamiltonian,

$$H_\mathrm{I} = H_\mathrm{Z} + H_Q, \qquad (1)$$

where $H_\mathrm{Z} = h\gamma \mathbf{I} \cdot \mathbf{B}_\mathrm{loc}$ is the Zeeman interaction expressed with the Planck constant $h$, the gyromagnetic ratio $\gamma = 11.199$ MHz/T, the nuclear spin $I = 7/2$ for the [51]V nucleus, and the local magnetic field $\mathbf{B}_\mathrm{loc}$, and $H_Q$ is the quadrupole interaction. In GaV$_4$Se$_8$, the effect of the second term $H_Q$ is too small to split the [51]V NMR peak. The resonance frequency of the [51]V nucleus $f_\mathrm{res}$ in GaV$_4$Se$_8$ is well approximated by $f_\mathrm{res} = \gamma |\mathbf{B}_\mathrm{loc}|$.

The local magnetic field $\mathbf{B}_\mathrm{loc}$ is composed of an external magnetic field $\mathbf{B}_\mathrm{ext}$, a hyperfine field $\mathbf{B}_\mathrm{hf}$ produced by the electron spin at the $a_1$ molecular orbital of the V$_4$ tetrahedron, classical dipole fields provided from electron spins at surrounding V ions, the Lorentz cavity field, and the demagnetization field. Since the hyperfine field $\mathbf{B}_\mathrm{hf}$ at the [51]V nucleus is much larger than the other contributions in GaV$_4$Se$_8$, we assume that the local magnetic field is approximated by the summation of the external field and the hyperfine field,

$$\mathbf{B}_\mathrm{loc} \cong \mathbf{B}_\mathrm{ext} + \mathbf{B}_\mathrm{hf}. \qquad (2)$$

The hyperfine field $\mathbf{B}_{\mathrm{hf}}$ is expressed as

$$\mathbf{B}_{\mathrm{hf}} = \mathbf{A}\boldsymbol{\mu}, \tag{3}$$

where $\mathbf{A}$ is the hyperfine coupling tensor and $\boldsymbol{\mu}$ is the magnetic moment at the V$_4$ tetrahedron.

Components of the coupling tensor are constrained by the local symmetry of the $^{51}$V site. Above $T_\mathrm{S} = 41$ K, the $^{51}$V nuclei are positioned at a single crystallographic site with a local symmetry of $.3m$. The coupling tensor is uniaxial along the local three-fold axis and characterized by two parameters $A_\perp$ and $A_\parallel$. Below $T_\mathrm{S}$, the $^{51}$V nuclei are positioned at two crystallographically inequivalent sites V1 and V2 with site symmetries of $3m$ and $.m$ respectively, which form an elongated V$_4$ tetrahedron, as shown in Fig. 1(b). Since the coupling tensor for the V1 site is also uniaxial along the polar axis, it can be expressed with two parameters $A_{1\perp}$ and $A_{1\parallel}$ in the local coordinate system (see Fig. 1(b) and Table I),

$$\mathbf{A_1} = \begin{pmatrix} A_{1\perp} & & \\ & A_{1\perp} & \\ & & A_{1\parallel} \end{pmatrix}. \tag{4}$$

In the same coordinate the coupling tensor $\mathbf{A}_2$ for the V2 site can be expressed with four parameters $A_{2\perp}$, $A'_{2\perp}$, $A_{2\parallel}$, and $\delta A_2$,

$$\mathbf{A_2} = \begin{pmatrix} A_{2\perp} & & \delta A_2 \\ & A'_{2\perp} & \\ \delta A_2 & & A_{2\parallel} \end{pmatrix}. \tag{5}$$

Here, note that the $y$ axis is perpendicular to the local mirror plane.

In the paramagnetic phase, we observed the $^{51}$V NMR spectra assigned to the V2 sites (see Appendix B). The spectra show the angle dependence of the resonance field with respect to the magnetic-field direction (Fig. B1), which can be explained by a nearly uniaxial NMR shift tensor characterized by two parameters $K_\mathrm{iso} = (2K_\perp + K_\parallel)/3$ and $K_\mathrm{ax} = (K_\parallel - K_\perp)/2$. This experimental result indicates that the coupling tensor $\mathbf{A}_2$ is nearly uniaxial, i.e. $A'_{2\perp} \sim A_{2\perp}$ and $\delta A_2 \sim 0$, which is probably because the $a_1$ molecular orbital occupied by an unpaired electron is constructed by the nearly uniaxial charge distributions around the V1 and V2 sites (see Fig. 1 (c)) [27,28]. However, it should be noted that the off-diagonal component $\delta A_2$ is tiny but not zero, which gives rise to a splitting of the NMR spectrum in the magnetically ordered phase under $\mathbf{B}_\mathrm{ext} \parallel$ [111]. This splitting allows us to assign the NMR spectrum to the V2 site in the ordered phase as shown later (see Discussions, Sec. IV A). The coupling constants for the $^{51}$V nuclei at the V1 and V2 sites $A_{1\perp}$, $A_{1\parallel}$, $A_{2\perp}$, $A_{2\parallel}$ are evaluated from analysis on the NMR shift versus magnetic susceptibility in the paramagnetic phase and magnetic-field dependence of the resonance frequencies as shown in Appendixes B and C. These parameters are listed in Table II. The large difference between $A_{1\perp}$ and $A_{2\perp}$ implies that the $3d$ orbital at the V1 site more significantly contributes to the $a_1$ molecular orbital than that at each V2 site.

Table II. Hyperfine coupling constants in the rhombohedral phase.

| Site | Component | Coupling constant (T/$\mu_B$) |
|---|---|---|
| V1 | $A_{1\perp}$ | $-8.38$ |
| V1 | $A_{1\parallel}$ | $-1.38$ |
| V2 | $A_{2\perp}$ | $-3.47$ |
| V2 | $A_{2\parallel}$ | $-0.13$ |

**III. Experimental Results**

**A. $^{51}$V NMR spectra at zero external magnetic field**

In the magnetically ordered phase, $^{51}$V NMR spectra are also observed at zero external magnetic field, because the ordered magnetic moment at the V ion produces a large internal field at the $^{51}$V nucleus. Figures 2(a) and 2(b) show the $^{51}$V NMR spectra observed at 4.2 K at zero external magnetic field. In the frequency range from 15 to 160 MHz, two asymmetric peaks are observed at ~30.7 and 73.7 MHz. As shown later (Sec. IV A), the high(low)-frequency peak denoted as $\beta$ ($\alpha$) is assigned to the V1 (V2) site (see Fig. 1(b)). Although the asymmetric shapes of these peaks are consistent with high-frequency edges of double-horn spectra expected for the incommensurate cycloidal structure at zero magnetic field [19], the continuous spectrum intensity which extends to lower frequencies to form the double-horn (see Fig. 6(a)), was not observed, implying a reduction of the NMR signal at lower frequencies. To clarify this reduction, the frequency dependence of the spin-echo decay rate $1/T_2$ was investigated around the high frequency peak. We found that $1/T_2$ is enhanced as the frequency decreases, as shown in the inset of Fig. 2(b), preventing us from observing the whole spectrum reflecting the accurate histogram of the resonance frequency. This frequency-dependent $1/T_2$ cannot be explained by nuclear dipole-dipole interactions [26], because the nuclear dipole contribution to the decay rate roughly estimated as $\gamma^2 h/r^3 \sim 3 \times 10^2$ s$^{-1}$ is much smaller than the observed large echo decay rate. Here, note that $r \sim 3 \times 10^{-8}$ cm is a distance between nearest-neighboring $^{51}$V nuclear spins. Instead, the decay rate is supposed to be dominantly governed by electron spin fluctuations. Since the lower-frequency part of the spectrum comes from the V sites with the magnetic moments inclined to the polar axis as shown in Sec. IV B, this frequency dependence of $1/T_2$ indicates the presence of an unconventional fluctuation of the spin component parallel to the polar axis, such as a phason [29,30], in the incommensurate cycloidal structure.

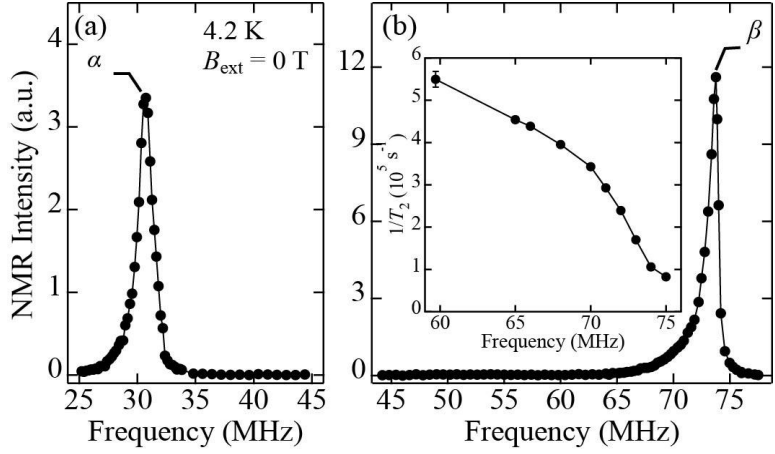

Fig. 2 (a, b) $^{51}$V NMR spectra observed at 4.2 K at zero external magnetic field. The inset of (b) shows the frequency dependence of the spin-echo decay rate $1/T_2$.

**B. Magnetization measurements at 4.2 K**

Figure 3 (a) shows the magnetic-field dependence of the magnetization $M$ measured at 4.2 K in the magnetic fields $\mathbf{B}_{\text{ext}}$ applied along the [001] and [111] axes. Both the magnetizations were measured in a field-increasing process. As shown in Fig. 3(a), the magnetization in $\mathbf{B}_{\text{ext}} \parallel [111]$ exhibits three successive metamagnetic transitions at $B_{\text{SkL}} = 0.08$ T, $B'_{\text{sat}} = 0.15$ T, and $B_{\text{sat}} = 0.38$ T, which can be clearly seen in the field derivative of the magnetization (Fig. 3(b)). As discussed in Ref. [19], two peaks in $dM/dB_{\text{ext}}$ at $B_{\text{SkL}}$ and $B_{\text{sat}}$ show the two magnetic phase transitions in domain A from the cycloidal state to the SkL state and the SkL state to the FM state, respectively. The step in $dM/dB_{\text{ext}}$ at $B'_{\text{sat}}$ shows the transition from the conical state to the FM state in the B–D domains (see Fig. 3(c)).

For $\mathbf{B}_{\text{ext}} \parallel [001]$, the magnetic field is tilted 54.7° from the polar axis of each domain A–D. Therefore, the direction of the magnetic field is equivalent for all four structural domains A-D, realizing the same magnetic state in all the domains. As shown in Fig. 3(a), the field dependence of the magnetization measured in $\mathbf{B}_{\text{ext}} \parallel [001]$ shows two kinks at $B^* = 0.08$ T and $B^*_{\text{sat}} = 0.16$ T, which are detected as a peak and a step in the $dM/dB_{\text{ext}}$ respectively. The step of $dM/dB_{\text{ext}}$ at $B^*_{\text{sat}}$ shows the phase transition to the FM state, whereas the origin of the peak at $B^*$ is not uniquely identified. The previous magnetization measurements conducted for various orientations of the magnetic field [17] show the persistence of the peak of $dM/dB_{\text{ext}}$ at $B^*$ by changing the direction of the external magnetic field from the [111] to the [001] axis, suggesting that the SkL state observed in $\mathbf{B}_{\text{ext}} \parallel [111]$ persists in oblique fields inclined to [001] axis (Fig. 3(d)). On the other hand, the torque magnetometry and the small-angle neutron scattering (SANS) measurements have demonstrated that the cycloidal state immediately turns into the conical state in $\mathbf{B}_{\text{ext}} \parallel [001]$, which persists up to the FM state without the SkL phase [23,24]. In this case, according to the ac susceptibility measurements, the peak of $dM/dB_{\text{ext}}$ at $B^*$ is suggested to be caused by a transformation of the magnetic structure in the structural domain walls to the ferromagnetic state [23].

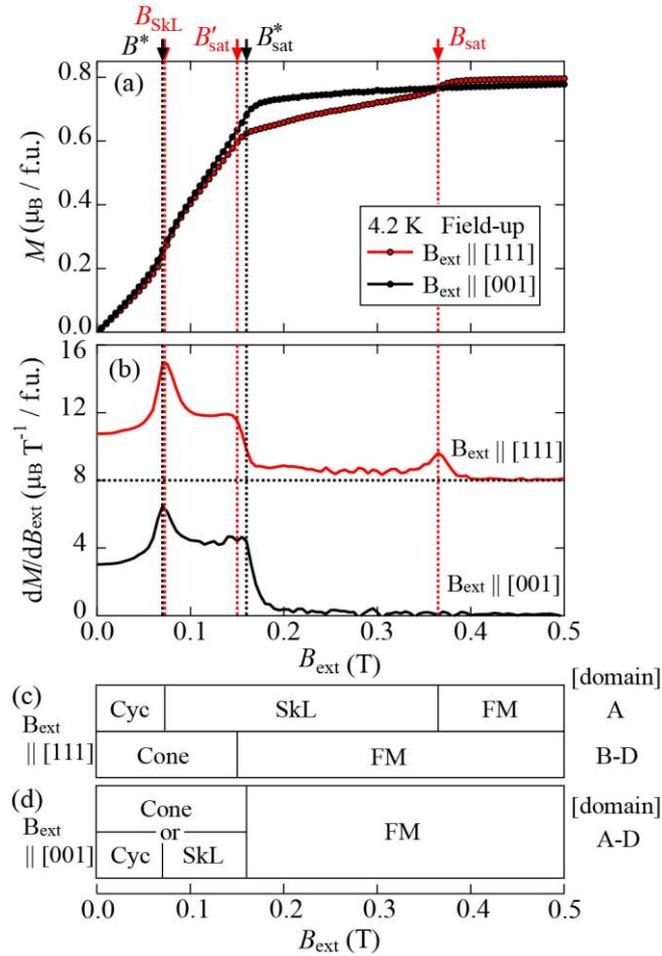

Fig. 3 Magnetic-field dependence of (a) magnetization and (b) its field derivative measured in the fields $\mathbf{B}_{\text{ext}} \parallel [111]$ (red) and $\mathbf{B}_{\text{ext}} \parallel [001]$ (black) at 4.2 K. Magnetic phase diagrams for the structural domain A and domains B-D in the external magnetic field (c) $\mathbf{B}_{\text{ext}} \parallel [111]$ and (d) two possible assignments in $\mathbf{B}_{\text{ext}} \parallel [001]$.

## C. $^{51}$V NMR spectra in $B_{ext} \parallel [111]$

Figure 4(a) shows the overall $^{51}$V NMR spectra observed at 4.2 K in various magnetic fields applied along the [111] axis. The NMR spectrum at each field is normalized by the intensity of the high-frequency peak for clarity. As shown in Fig. 4(a), both the peaks $\alpha$ and $\beta$ observed at 30.7 and 73.7 MHz at zero external magnetic field shift to lower frequencies with increasing $B_{ext}$ with changes in their shapes. The peak $\alpha$ splits into two ($\alpha1$ and $\alpha2$) above 0.15 T (see Fig. 4(b)). As the peak $\beta1$ shifts to the lower frequency at higher $B_{ext}$, another absorption peak $\gamma$ becomes discernible above 0.13 T at around 73.7 MHz. The frequency of the peak $\gamma$ does not depend on the applied field, which disappears above 0.2 T. Besides these peaks, additional broad peaks (denoted as $\beta2$ and $\beta3$) become discernable above 0.10 T at around 50 MHz, which shift to the higher frequency with increasing $B_{ext}$ and merge with $\beta1$ at around 0.17 T.

To further investigate the split peak observed at the high-frequency edge in more detail, we measured $^{51}$V NMR at a constant magnetic field separately in the magnetization process up to 4 T and in the demagnetization process after the magnetization process (Fig. 5). As shown in Fig. 5, the peak $\gamma$ observed up to ~0.2 T in the magnetization process completely disappears in the demagnetization process, showing a magnetic hysteresis of this absorption signal.

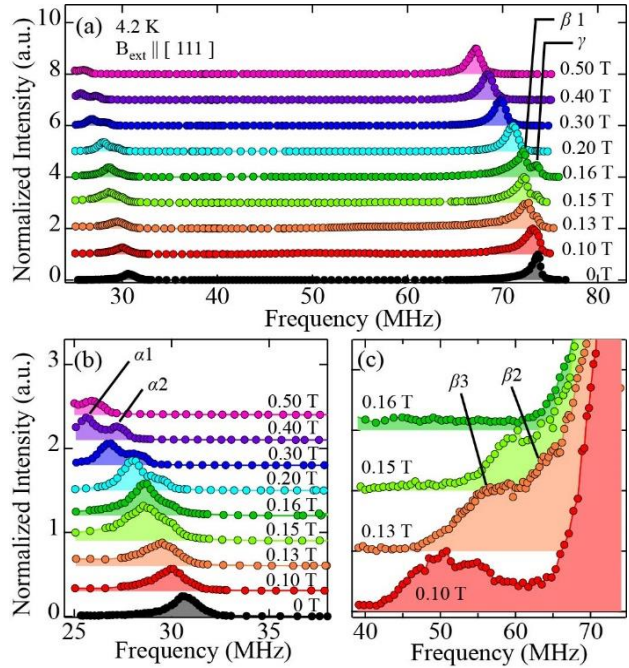

Fig. 4 (a) $^{51}$V NMR spectra observed in various external magnetic fields applied along the [111] axis at 4.2 K. The enlarged NMR spectra around 32 and 55 MHz are shown in (b) and (c), respectively. The spectra are plotted with offsets.

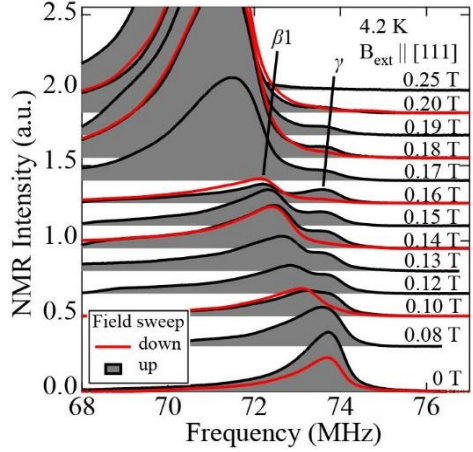

Fig. 5 $^{51}$V NMR spectra at 4.2 K measured in the magnetization (grey) and the demagnetization (red line) processes of the external magnetic field applied along the [111] axis. The spectra are plotted with offsets.

## D. $^{51}$V NMR spectra in $\mathbf{B}_{ext} \parallel [001]$

Figure 6 (a) shows the overall $^{51}$V NMR spectra observed at 4.2 K in various magnetic fields applied along the [001] axis. Similar to the spectra observed in $\mathbf{B}_{ext} \parallel [111]$, the two peaks observed at zero external magnetic field shift to the lower frequencies, whereas the spectral shapes differently change. As shown in Figs. 6(b) and (c), new peaks $\alpha 3$, $\beta 4$, and $\beta 5$ become discernible above ~0.12 T at the lower-frequency side of both the two peaks at zero field. These new peaks shift to the higher frequencies at higher magnetic field until they merge with peaks $\alpha 4$ and $\beta 1$, which are smoothly connected to the two peaks at zero field. This convergence of the spectral shape to the single peak toward 0.17 T is caused by the magnetic phase transition to the FM state.

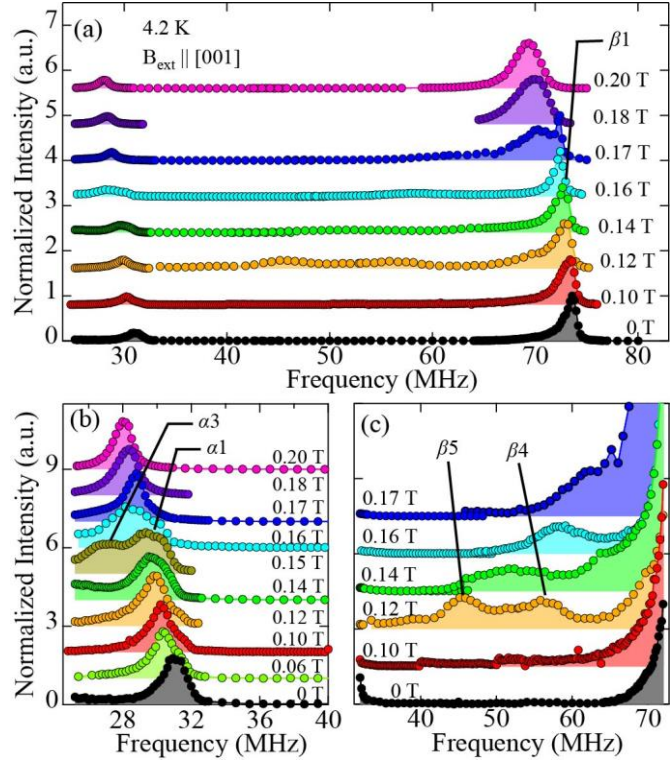

Fig. 6 (a) $^{51}$V NMR spectra observed in various external magnetic fields applied along the [001] axis at 4.2 K. The enlarged NMR spectra around 32 and 55 MHz are shown in (b) and (c), respectively. The spectra are plotted with offsets.

## IV. Discussions
### A. Site assignments of the NMR spectrum for $B_{ext} \parallel [111]$

For site-assignment of the observed absorption peaks for $B_{ext} \parallel [111]$, let us first focus on the FM state above 0.4 T. In the FM state four sets of $^{51}$V NMR peaks are supposed to be observed, which are assigned to the two crystallographically inequivalent sites V1 and V2 in two kinds of structural domain, domain A and domains B-D. For $B_{ext} \parallel [111]$, the magnetic field is parallel to the polar axis of one of the structural domain (domain A), while it is tilted by 109.5° from the polar axes of domain B-D. The spectrum for domain A has a single peak for each of the V1 and V2 sites, which shifts linearly with increasing magnetic field. This is because the ordered magnetic moment is (nearly) parallel to the principal axis of the hyperfine coupling tensor for the $^{51}$V nucleus at the V1 (V2) site. In domains B-D the spectrum for the V1 (V2) site is supposed to have a single (double) peak, because the hyperfine coupling tensor for the V1 site is uniaxial around the polar axis due to the local symmetry whereas that for the V2 is not (see Eq.(5)). The resonance frequencies of these peaks depend on the external magnetic field nonlinearly, because the magnetic field is neither parallel nor perpendicular to the principal axis of the hyperfine coupling tensor.

At 0.4 T, we observed three absorption peaks $\alpha 1$, $\alpha 2$, and $\beta 1$ as shown in Figs. 4(a) and 4(b). The resonance frequencies for these three absorption peaks show magnetic-field dependence as shown in Fig. 7(a). From the nonlinear magnetic field dependence of these resonance frequencies, the peak $\beta 1$ and the doublet peaks $\alpha 1$ and $\alpha 2$ can be assigned to the V1 and V2 sites of domains B-D, respectively. Indeed, the resonance frequencies of these peaks can be nearly reproduced by the calculated frequency $f_{\text{res}} = \gamma |\mathbf{B}_{\text{ext}} + \mathbf{A}_i \boldsymbol{\mu}|$ with the coupling constants $A_{i\perp}$ and $A_{i\parallel}$ ($i = 1,2$) listed in Table II as shown by the solid black and blue lines in Fig. 7(a). Note that the data points of peak $\beta 1$ deviate from the calculated line for the V1 site below 2 T, because the ordered magnetic moment is inclined from the field direction (see Sec. IV C). The doublet peaks are caused by tiny off-diagonal components of the hyperfine coupling tensor for the V2 site.

Besides the three absorption peaks, another peak $\alpha 5$ is observed in the FM state above 4 T as shown in Fig. 7(b). The peak $\alpha 5$ shifts linearly with increasing magnetic field as shown in Fig. 7 (a), which can be reproduced by the calculated frequency for the V2 site in domain A. Although the absorption peak for the V1 site in domain A should be detected at the resonance frequency $f_{\text{res}}(B_{\text{ext}})$ indicated by the black dashed line in Fig. 7(a), we did not detect this absorption peak probably due to a short spin-echo decay time.

Next, let us focus on the spectra observed at lower magnetic fields less than 0.4 T. The peak $\beta 1$ and the doublet peaks $\alpha 1$ and $\alpha 2$ are smoothly connected to the peaks at 73.7 and 30.7 MHz in zero external magnetic field, respectively, which indicates that the peak at 73.7 MHz (30.7 MHz) is assigned to the V1 (V2) site. The broad doublet peaks $\beta 2$ and $\beta 3$ at around 50 MHz discernable above 0.1 T are assigned to the low-frequency edge of the spectra of the conical state of the V1 site in the domains B-D, since it converges to the peak $\beta 1$ at higher magnetic fields as shown later (Fig. 10). The remaining peak $\gamma$ observed below 0.2 T at the constant frequency of 73.7 MHz is assigned to the V1 site in domain A, since all the other absorption peaks from domains B-D are assigned. The resonance frequency of the peak $\gamma$ is not continuously connected to the signals in the FM state, because of the discontinuous change of the magnetic structure from the SkL to the FM state in domain A, in contrast to the continuous closing of the cone state in domains B-D. The NMR signal of the V2 sites in domain A expected at around 30 MHz is not identified because it overlaps with the splitting of the V2 signal (the doublet peaks $\alpha 1$ and $\alpha 2$) in domains B-D at around 30 MHz.

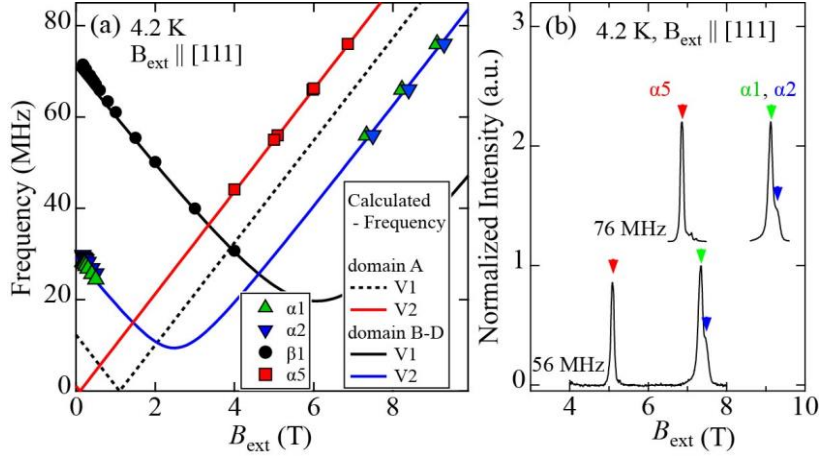

Fig. 7 (a) Magnetic-field dependence of the resonance frequencies for the $^{51}$V NMR absorption peaks in the external magnetic field $\mathbf{B}_\text{ext} \parallel [111]$. The solid and dashed lines are calculated resonance frequencies with an assumption that the ordered magnetic moments are parallel to the magnetic field and the coupling constants follow the relations $A_{1\perp} = -8.38\ \text{T}/\mu_B$ and $A_{1\parallel} = -1.38\ \text{T}/\mu_B$ for the $^{51}$V nucleus at the V1 site and $A_{2\perp} = -3.47\ \text{T}/\mu_B$ and $A_{2\parallel} = -0.13\ \text{T}/\mu_B$ for the $^{51}$V at the V2 site. (b) $^{51}$V NMR spectra taken by field-sweep measurements at fixed frequencies of 56 and 76 MHz. The spectrum for 76 MHz is plotted with an offset.

## B. Calculation of the $^{51}$V NMR spectrum in the magnetically ordered phase at zero external magnetic field

In general, an NMR spectrum in a magnetically ordered phase is nothing but a histogram of the distributed local magnetic fields. In the case of GaV$_4$Se$_8$, the local magnetic field at a $^{51}$V nucleus is composed of the external magnetic field and the hyperfine field given by Eq. (3). In this section, we demonstrate calculated histograms of the NMR frequencies for cycloidal and Néel-type skyrmion lattice states with incommensurate propagation vectors.

Let us focus on a $^{51}$V nucleus at a V1 site in zero external magnetic field. The NMR frequency of the $^{51}$V nucleus in the magnetically ordered phase can be expressed by $f_\text{res} = \gamma |\mathbf{B}_\text{loc}|$, where $\mathbf{B}_\text{loc} = \mathbf{A}_1 \boldsymbol{\mu}$. For an incommensurate cycloidal state with the cycloidal plane perpendicular to the $Y$ axis in the local coordinate system, hyperfine field $\mathbf{B}_\text{cyc}$ for a $^{51}$V nucleus at the V1 site can be expressed as

$$\mathbf{B}_\text{cyc} = \begin{pmatrix} A_{1\perp} & & \\ & A_{1\perp} & \\ & & A_{1\parallel} \end{pmatrix} \begin{pmatrix} \mu \sin\theta \\ 0 \\ \mu \cos\theta \end{pmatrix}, \qquad (6)$$

where $\mu$ is the magnitude of the ordered magnetic moment at a V$_4$ tetrahedron and the polar angle $\theta$ is an angle between the polar axis $Z$ and the ordered moment. In zero external magnetic field the magnetic moment $\mu$ is determined as $\mu = 0.79$ $\mu_B$ to reproduce the resonance frequency of the observed spectrum by our calculation (Fig. 8(a)). Since the angle $\theta$ is distributed from 0 to 180° uniformly in the incommensurate cycloidal state and the coupling tensor $\mathbf{A}_1$ has a large anisotropy of $A_{1\perp} = -8.38$ T/$\mu_B$ and $A_{1\parallel} = -1.38$ T/$\mu_B$ as shown in Table II, the histogram of the resonance frequency at zero external magnetic field is broadly distributed in the frequency range between 12.1 and 73.7 MHz. As shown in Fig. 8(a), the histogram has two peaks at the low- and high-frequency edges. While the low-frequency peak is contributed from the V1 sites with the ordered magnetic moments pointing to the polar axis ($\theta = 0°, 180°$), the high-frequency peak is formed by the V1 sites with the moments perpendicular to the polar axis ($\theta = 90°, 270°$). Thus, the resonance frequency has one-to-one correspondence to the polar angle of the spin orientation around the polar axis due to the uniaxial coupling tensor. For the V2 site, it is expected that the histogram of the resonance frequency also has a double-horn shape in the cycloidal state. Since the hyperfine coupling constant $A_{2\perp}$ is smaller than $A_{1\perp}$ (see Table II), the high-frequency peak appears at a lower frequency of 30.7 MHz, which is observed as a peak α at 4.2 K as shown in Fig. 2 (a).

As is the case for the cycloidal state, the histogram of the $^{51}$V NMR frequency for the Néel-type SkL is obtained by considering the spin alignments on the crystal lattice (see Appendix D). Figure 8(b) shows the histogram for the Néel-type SkL state. The calculated histogram has the peak and the step at the high- and low-frequency edges, respectively, as well as the small peak at around 25 MHz. A distinct difference between the histograms for the cycloidal and SkL states is the location of this low-frequency peak. The small low-frequency peak for the SkL state is formed by the V1 sites located in the core and the peripheral parts of the magnetic skyrmions where the ordered magnetic moments are almost parallel to the polar axis as shown in Fig. 8(c).

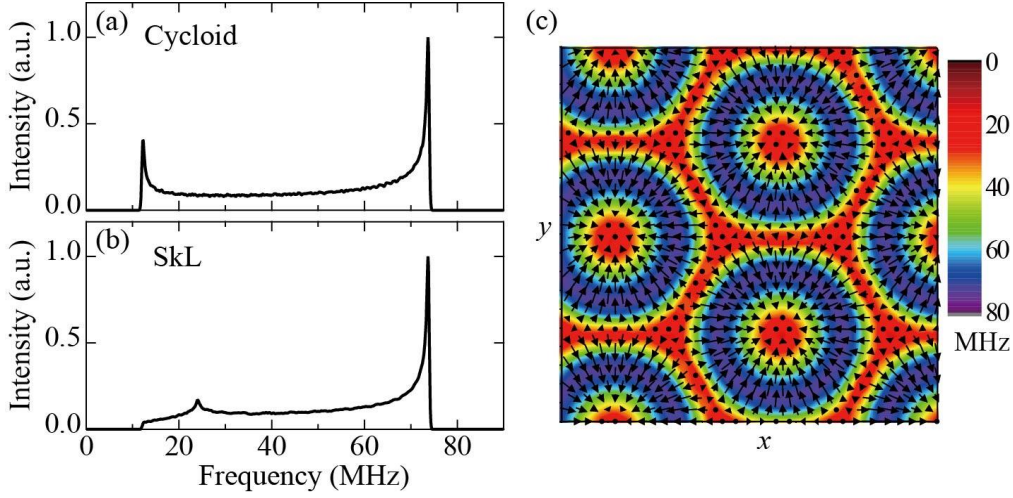

Fig. 8 Calculated histogram of the $^{51}$V NMR frequencies for the incommensurate (a) cycloidal and (b) the skyrmion lattice states in GaV$_4$Se$_8$ at zero external magnetic field. (c) Distribution of the NMR frequency (color scale) and the projected magnetic moments (black arrows) in the (111) plane for the skyrmion lattice state.

### C. Magnetic structure in $\mathbf{B}_{\text{ext}} \parallel [111]$

In order to discuss magnetic-field dependence of the magnetic structure in $\mathbf{B}_{\text{ext}} \parallel [111]$, we focus on the $^{51}$V NMR absorption peaks assigned to the V1 site. As mentioned in the Sec. IV A, the $^{51}$V NMR peaks from the V1 site assigned to domain A (peak $\gamma$) and domains B-D (peaks $\beta$1, $\beta$2 and $\beta$3) are observed separately.

First, we focus on peaks $\beta$1, $\beta$2 and $\beta$3 assigned to the V1 sites in domain B-D. In domains B-D, the cycloidal state is immediately changed to the conical state by applying a magnetic field, which turns into a FM state above $B'_{\text{sat}}$ (Figs. 9(a)-9(c)). For instance, in domain C where the applied magnetic field is tilted by $19.5°$ from the local axis $X_3$ ($\parallel [112]$) in the $Z_3 X_3$ plane ($\perp [\bar{1}10]$) (see Figs. 9(d), 9(e)), a cycloidal state with a single-$q$ vector parallel to the $Y_3$ ($\parallel [\bar{1}10]$) axis is selected by applying a small magnetic field of ~0.05 T [19] (see Fig. 9(a)). As the magnetic field increases, the cycloid is turned into the conical state which closes with the cone axis tilting from the $X_3$ axis to $[111]$ as shown in Figs. 9(b) and 9(c). In the local coordinate system for domain C (see Table I), the ordered magnetic moments on the cone can be expressed by

$$\boldsymbol{\mu}_C = \mu \begin{pmatrix} \cos\xi \cos\zeta + \sin\xi \sin\zeta \cos\phi \\ \sin\xi \sin\phi \\ -\cos\xi \sin\zeta + \sin\xi \cos\zeta \cos\phi \end{pmatrix}, \quad (7)$$

where $\phi$ is the azimuth angle, and $\xi$ and $\zeta$ are the opening and tilting angles of the cone, respectively as shown in Figs. 9(d) and 9(e). The azimuth angle is uniformly distributed over all the angle range.

From this spin distribution we can obtain the histogram of the resonance frequency for the V1 sites in domain C. Note that this histogram is identical to those for domain B and D, since the orientation of the external magnetic field is equivalent for these three domains. For a set of angles $\xi$ and $\zeta$, we obtain a histogram of the resonance frequency $f_{res} = \gamma|\mathbf{B}_{ext} + \mathbf{A}_1\boldsymbol{\mu}_C|$ with three peaks as shown in Figs. 10(a) and 10(b). The peak at the highest frequency comes from the ordered moment with $\phi = 90°$ and $270°$, while the peaks at the lowest and middle frequencies are from those with $\phi = 0°$ and $\phi = 180°$ respectively. The resonance frequencies of the three peaks depend on the angles $\xi$ and $\zeta$. For instance, the interval frequency between the two peaks at the highest and lowest frequencies decreases as the opening angle $\xi$ degreases from $90°$.

Although the observed NMR spectrum cannot be exactly reproduced by the calculated histogram, especially for the low-frequency peaks, convergence of the two peaks $\beta 2$ and $\beta 3$ to the peak $\beta 1$ can be explained by the calculated histogram of the NMR frequencies with assuming that the angles $\xi$ and $\zeta$ exhibit magnetic-field dependence as shown in Fig. 9(f). As the applied field $B_{ext}$ increases, the opening angle $\xi$ decreases toward zero up to 0.17 T, whereas the angle $\zeta$ keeps increasing above 0.17 T, which indicates that the ordered magnetic moments do not point to the magnetic-field orientation just after the cone closes at $B'_{sat}$. Instead, it is required to apply the external magnetic field above 3 T to make the orientation of the ordered moments parallel to the applied magnetic field ($\zeta = 19.5°$) due to the strong in-plane anisotropy of the magnetic moments.

Next, let us turn to the peak $\gamma$ assigned to the V1 site in domain A. In domain A, the cycloidal state at zero external field changes to the SkL state above $B_{SkL} = 0.08$ T. Assuming that the magnetic structures of both the cycloidal and the SkL states are not changed by an applied magnetic field, we calculate the field dependence of the histograms of the resonance frequency as shown in Figs. 10(a) and 10(b). As shown in Figs. 10(a) and 10(b), the calculated histograms of these two states are almost identical above 40 MHz. Both histograms have an asymmetric peak at the high-frequency edge which is hardly moved by increasing the applied magnetic field. This is because the high-frequency peaks for both states come from the V1 sites in which the magnetic moments are perpendicular to the polar axis. Given that the peak $\gamma$ remains at the high-frequency edge in the magnetic field up to 0.2 T, the peak $\gamma$ can be assigned to the V1 sites with the magnetic moments perpendicular to the polar axis in either cycloidal or SkL states.

Although the peak $\gamma$ was resolved in the SkL state above $B_{SkL}$, the peak $\gamma$ disappears in the SkL state for $0.2~\text{T} < B_{\text{ext}} < B_{\text{sat}}$ and was not observed at all in the SkL state in the demagnetization process (Fig. 5). These results suggest that the peak $\gamma$ does not come from the SkL state but from the supermagnetized cycloidal state that persists above $B_{\text{SkL}}$ only in the magnetization process due to the first-order transition nature.

This magnetic hysteresis caused by the supermagnetized cycloidal state is consistent with the magnetic hysteresis observed in the thermal Hall conductivity ($\kappa_{xy}$) in GaV$_4$Se$_8$ [21]. According to Ref. [21], the magnitude of $\kappa_{xy}$ reflects the volume fraction of the SkL state, which shows a sharp increase (decrease) as it enters (exists from) the SkL state at 10 K. However, at 5 K, the increase of $\kappa_{xy}$ from the cycloidal to the SkL state in the magnetization process becomes gradual up to around 0.25 T whereas that in the demagnetization one remains sharp. This gradual increase of $\kappa_{xy}$ in the magnetization process can be explained by a gradual decrease of the supermagnetized cycloidal state, and the sharp increase of $\kappa_{xy}$ in the demagnetization process is due to the absence of the supermagnetized state. Thus, the magnetic hysteresis observed in the $^{51}$V NMR spectrum and the thermal Hall conductivity $\kappa_{xy}$ suggests the presence of the supercooled cycloidal state in the magnetization process below 5 K.

The site assignment of the peak $\gamma$ to the cycloidal state indicates the absence of the NMR signal from the SkL state. At the end of this section, let us briefly comment on the absence of the NMR signal from the SkL state. We suggest that the NMR signal from the SkL state was missing because the spin-echo decay time $T_2$ for the $^{51}$V nuclei in the SkL state is too short to detect the spin-echo signal as reported in other SkL host material [31]. This short $T_2$ in the SkL state might be caused by additional low-energy excitations allowed only in the SkL state, such as the clockwise and/or counter-clockwise motions or the breathing modes of the magnetic skyrmions as observed in GaV$_4$S$_8$ [32], because $T_2$ is determined by slow spin fluctuations [26]. NMR on a different ligand site with a weaker hyperfine coupling may be advantageous for detecting the signals from the SkL state because $T_2$ is increased by the weaker coupling. In the case of GaV$_4$Se$_8$, $^{69}$Ga, $^{71}$Ga and $^{77}$Se nuclei are possible candidates. However, NMR measurements of these nuclei would be difficult due to lower sensitivity by the lower NMR frequency or smaller internal fields, which remains as a future issue.

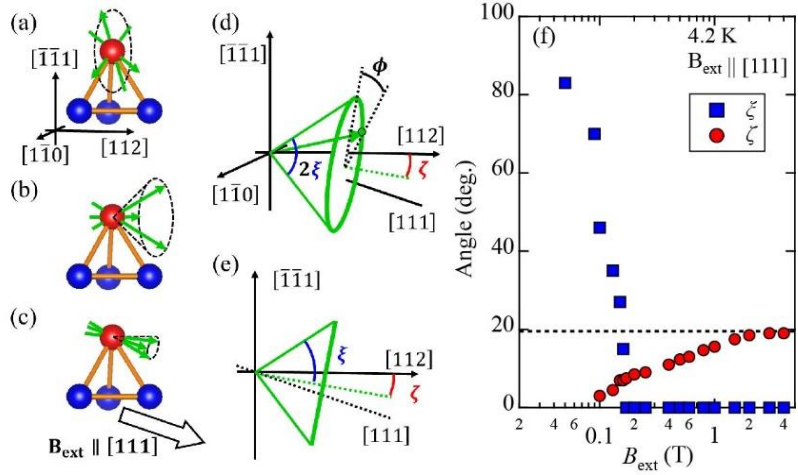

Fig. 9 Schematic figures of the magnetic-field evolution of the distributed magnetic moments at the V1 site in the structural domain C under (a) the zero, (b) a medium, and (c) an almost saturating applied magnetic field along the [111] direction. (d, e) Perspective and projected views of a cone formed by the distributed magnetic moments in the local orthogonal coordinate system of domain C. The green arrow in (d) represents a magnetic moment whose vector is characterized by an azimuth angle $\phi$, an opening angle $\xi$, and a tilting angle $\zeta$ of the cone. (f) Magnetic-field dependence of the opening and tilting angles, $\xi$ and $\zeta$ for the conical state. The dotted line in (f) shows angle $\zeta$ when the magnetic moment becomes parallel to the applied field.

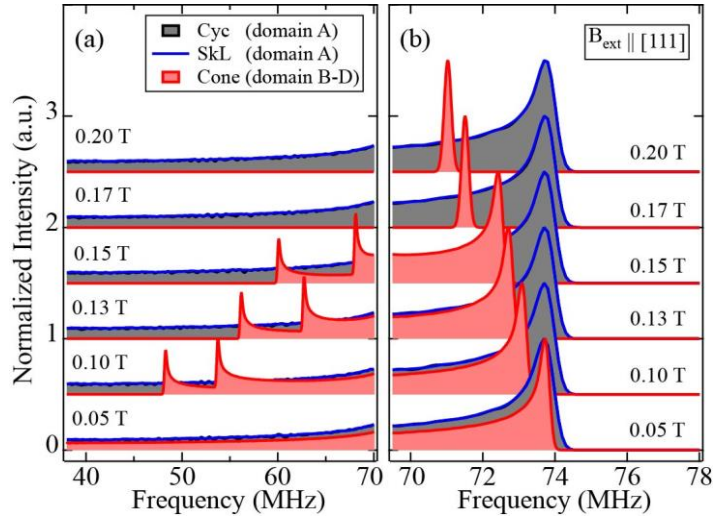

Fig. 10 Field dependence of the calculated histogram of the $^{51}$V NMR frequencies in $\mathbf{B}_{\text{ext}} \parallel [111]$ for the conical state in domains B–D (red), and those in the cycloidal (grey) and the SkL (blue) states in domain A. (a) The calculated histogram in the frequency range between 38 and 70 MHz, and (b) the enlarged high-frequency edge around 74 MHz. The histograms are plotted with offsets.

**D. Magnetic structure in $\mathbf{B}_{\text{ext}} \parallel [001]$**

In $\mathbf{B}_{\text{ext}} \parallel [001]$, as discussed in the Experimental Results (Sec. III B), the magnetic state for $0 < B_{\text{ext}} < B'_{\text{sat}}$ is suggested to be either the conical state or the SkL state transformed from the cycloidal state above $B^*$. We discuss these two suggestions by comparing our $^{51}$V NMR spectrum to the calculations of these two states in the oblique field.

First, let us demonstrate that the magnetic-field dependence of the observed spectrum can be reproduced by the transformation of the magnetic structure from the cycloidal state to a conical state as discussed in the previous section. To calculate the histogram of the resonance frequency, we focus on the magnetic structure in domain A. In the local coordinate system, an ordered magnetic moment can be expressed by

$$\mu_A = \mu \begin{pmatrix} \sin\xi \sin\phi \\ \cos\xi \cos\zeta - \sin\xi \sin\zeta \cos\phi \\ \cos\xi \sin\zeta + \sin\xi \cos\zeta \cos\phi \end{pmatrix}, \tag{8}$$

where $\phi$ is an azimuth angle distributed from 0° to 360° uniformly, and $\xi$ and $\zeta$ are the opening and tilting angles of the cone respectively as defined in $\mathbf{B}_{\text{ext}} \parallel [111]$. As in $\mathbf{B}_{\text{ext}} \parallel [111]$, the $^{51}$V NMR frequency given by $f_{\text{res}} = \gamma|\mathbf{B}_{\text{ext}} + \mathbf{A}_1 \boldsymbol{\mu}_A|$ is distributed for a set of angles $\xi$ and $\zeta$, which results in a histogram with three peaks as shown in Figs. 11(a) and 11(b). The resonance frequencies of these three peaks depend on the angles $\xi$ and $\zeta$. The three observed peaks, $\beta 1$, $\beta 4$, and $\beta 5$ can be assigned to the three peaks in the calculated histogram if we assume the magnetic-field dependence of the angles $\xi$ and $\zeta$ shown in Fig. 12. The opening angle of the cone smoothly approaches zero up to 0.18 T, while the tilting angle $\zeta$ saturates around 3 T due to the in-plane anisotropy of the ordered magnetic moment. Thus, the magnetic-field evolution of the NMR spectrum can be explained by the field dependence of the conical state.

Next, we show that the observed field dependence of the NMR spectrum is also consistent with the field dependence of the magnetic skyrmion lattice skewed in $\mathbf{B}_{\text{ext}} \parallel [001]$. Since the magnetic field applied along the [001] axis is tilted by 54.7° from the polar axis, the magnetic structure of the SkL is deformed from that in $\mathbf{B}_{\text{ext}} \parallel [111]$ due to the magnetic easy-plane anisotropy with respect to the polar axis. To clarify how the skyrmion is deformed in the oblique magnetic field, we performed micromagnetic calculations based on the Landau–Lifshitz–Gilbert (LLG) equation by using MuMax3 software package [33]. The calculations were conducted for a two-dimensional square lattice with $512 \times 512$ sites under several magnetic fields tilted 54.7° from the $z$ axis which is parallel to the polar axis. The magnetization distribution is deduced by minimizing the magnetic free energy $\int \epsilon(\mathbf{r}) d\mathbf{r}$ with energy density $\epsilon(\mathbf{r})$ expressed by the sum of the magnetic exchange, DM, Zeeman, and the anisotropy energy contributions [19],

$$\epsilon = A_{\text{ex}}(\nabla \mathbf{m})^2 + \epsilon_{DM} - M_{\text{sat}} \mathbf{m} \cdot \mathbf{B}_{\text{ext}} - k_u m_z^2, \quad (9)$$

where $A_{\text{ex}}$ is a magnetic exchange coupling constant, $\mathbf{m}$ is a reduced magnetization, $M_{\text{sat}}$ is a saturation magnetization, $k_u$ is a magnetic anisotropy coupling constant, and $\epsilon_{DM}$ is given by $\epsilon_{DM} = D(m_x \partial_x m_z - m_z \partial_x m_x + m_y \partial_y m_z - m_z \partial_y m_y)$ with a DM coupling constant $D$ due to $C_{3v}$ symmetry. In the simulation, we used the parameters $A_{\text{ex}} = 8.8 \times 10^{-14}$ J/m, $M_{\text{sat}} = 35.6$ kA/m, $K_u = -10$ kJ/m$^3$, and $D = 6.0 \times 10^{-5}$ J/m$^2$ and a Gilbert damping parameter for the LLG equation 0.1, taking into account lattice parameters [34], magnetic transition temperature $T_c = 18$ K, and the strength of DM interaction [20] of GaV$_4$Se$_8$. These parameters used in our simulation satisfy an effective anisotropy parameter $|A_{\text{ex}} k_u / D^2| \sim 0.24$, which indicates a moderate easy-plane anisotropy appropriate for GaV$_4$Se$_8$ [19]. These parameters are also comparable to those used in the micromagnetic simulation for an isostructural compound GaV$_4$S$_8$ [35]. In the case of GaV$_4$S$_8$, the simulation reproduces experimental results of the optical measurement [35], which justifies the parameters used in our calculations. In order to trace the deformation of the skyrmion, first we made a regular SkL state stabilized in zero external magnetic field by demagnetizing the external magnetic field after the SkL state was created in an external magnetic field applied along $z$ axis. Then, we investigated the magnetization distribution with changing the external magnetic field applied along the oblique direction.

Figures 13(a)-13(d) show the counter maps of the $z$ components of spin $S_z$ for the calculated skewed SkL on a square lattice with a size of $90 \times 90$ cells in external magnetic fields from 0 to $B_{\text{ext}}/B_{\text{sat}} = 0.55$, where the saturation field $B_{\text{sat}}$ in our calculation is determined by the field dependence of the calculated magnetization. As shown in Figs. 13(a)–13(d), the core of the magnetic skyrmions moves in the direction opposite to the in-plane field with increasing magnetic fields (Fig. 13(d)) from its central position at zero external field (Fig. 13(a)). From the calculated spin structures, the field dependence of the histogram of the NMR frequencies is calculated as shown in Figs. 14(a) and 14(b). As the magnetic field increases, the peak at the high-frequency edge shifts to the lower frequency, while the small peak observed at 25 MHz in zero external magnetic field shifts to the higher frequency. This magnetic-field evolution in the calculated histogram qualitatively reproduces the convergence of peaks $\beta 4$ and $\beta 5$ to the peak $\beta 1$ observed in $\mathbf{B}_{\text{ext}} \parallel [001]$ (Fig. 6(c)). In particular, peaks $\beta 4$ and $\beta 5$ discernable above 0.12 T in the experiment may be explained by splitting of the low-frequency peak of the calculated histogram at around $B_{\text{ext}}/B_{\text{sat}} = 0.55$ (Fig. 15(a)).

Let us focus on the calculated histogram at $B_{\text{ext}}/B_{\text{sat}} = 0.55$ showing the two humps at the low-frequency edge (red arrows in Fig. 14(a)). These two humps show that there are the two local maxima in the distribution of the direction of the magnetic moments, which are indicated by the red and the black regions in Fig. 15(b). As shown in Fig. 15(b), the V sites corresponding to the two humps are mostly located in the peripheral region between the skyrmions. To check the spin structure in these regions, the spin configurations in the cross section along the rectangles in Fig. 15(b) are shown in Fig. 15(c). As shown in Fig. 15(c), the directions of the magnetic moments in the red and the black regions are oriented to the special angles which would be determined by the competition between the Zeeman energy and the DM interaction in the skewed skyrmion lattice in the oblique field. Since the presence of the two humps is characteristic of the SkL in the oblique field, the peaks $\beta 4$ and $\beta 5$ observed in our measurements supports the appearance of the skewed SkL in $\mathbf{B}_{\text{ext}} \parallel [001]$.

As a result, the observed NMR spectrum can be explained by either of the two calculations of the conical state and the skewed SkL. This is because the difference in spectral shape at low frequencies between the two magnetic states, i.e. the distinct two peaks in the conical state (Fig. 11) and the continuous signal in the skewed SkL (Fig. 14(a)), was not detected in our measurements due to the damping of the spectral intensity at lower frequencies by larger spin-echo decay rate (Fig. 2). On the other hand, given the absence of the NMR signal in $\mathbf{B}_{\text{ext}} \parallel [111]$, it would be more likely that the observed peaks $\beta 1$, $\beta 4$, and $\beta 5$ in $\mathbf{B}_{\text{ext}} \parallel [001]$ (Fig. 6) are coming from the conical state rather than the skewed SkL. To verify the presence and/or absence of the SkL state in the oblique magnetic field, it is necessary to detect the NMR signals at lower frequencies or to use other experimental probe such thermal Hall measurements, which remains as a future work.

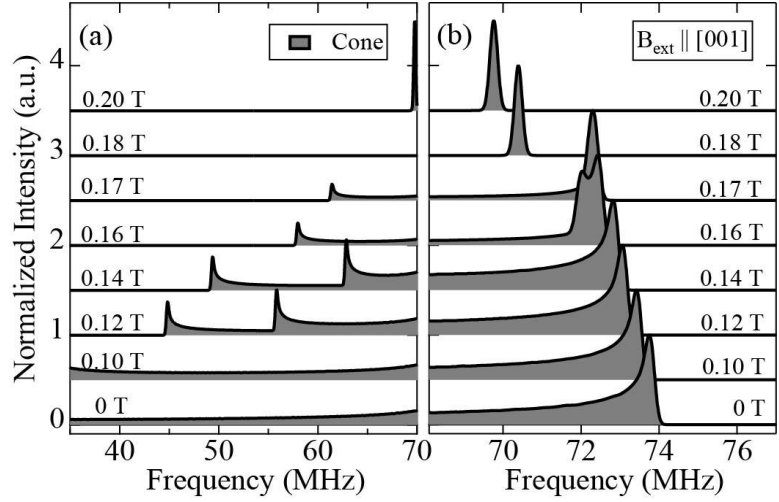

Fig. 11 Magnetic-field dependence of the calculated histogram of the $^{51}$V NMR frequencies of the conical state in $\mathbf{B}_{ext} \parallel [001]$. (a) Enlarged frequency range around 55 MHz, and (b) the high-frequency edges of the histogram. The histograms are plotted with offsets.

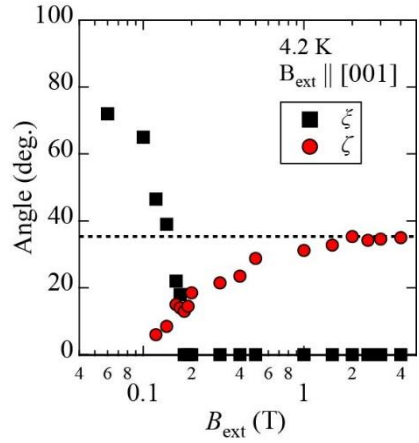

Fig. 12 Magnetic-field dependence of the opening and tilting angles $\xi$ and $\zeta$ of the conical state in $\mathbf{B}_{ext} \parallel [001]$. The dotted line shows the angle $\zeta$ when the magnetic moment becomes parallel to the applied field.

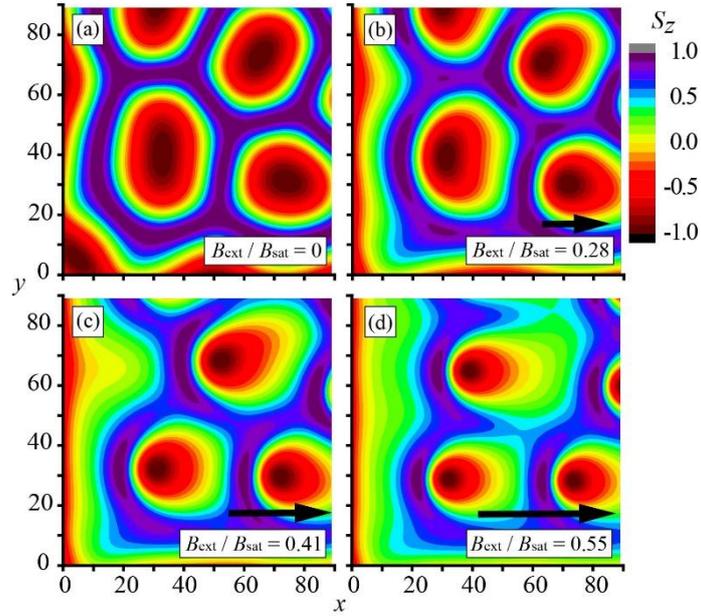

Fig. 13 Color maps of the polar $z$ component of the spins $S_z$ for the calculated skewed SkL on a square lattice with the size of $90 \times 90$ cells in $B_{ext}/B_{sat} =$ (a) 0, (b) 0.28, (c) 0.41, and (d) 0.55. The external magnetic field is tilted $54.7°$ from the $z$ axis to the $x$ axis as shown by the black arrow.

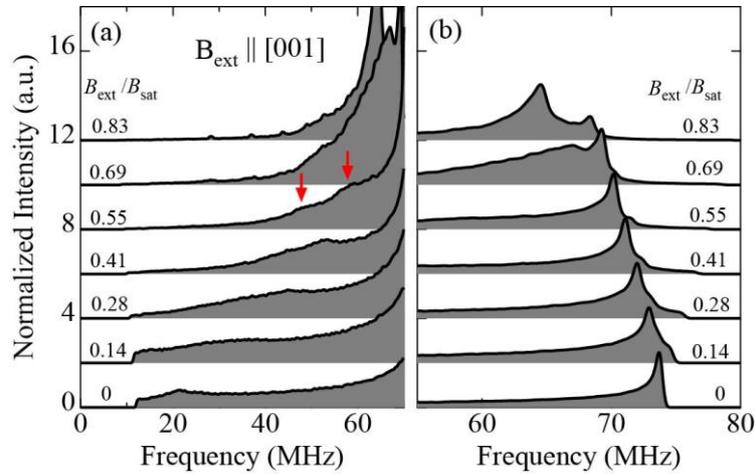

Fig. 14 Calculated histogram of the $^{51}$V NMR frequency of the skewed SkL state in $\mathbf{B}_{ext} \parallel [001]$ around (a) the low-frequency edge and (b) the high-frequency edge. The red arrows in (a) indicate two humps appearing at around the low frequency edge of the histogram. The histograms are plotted with offsets.

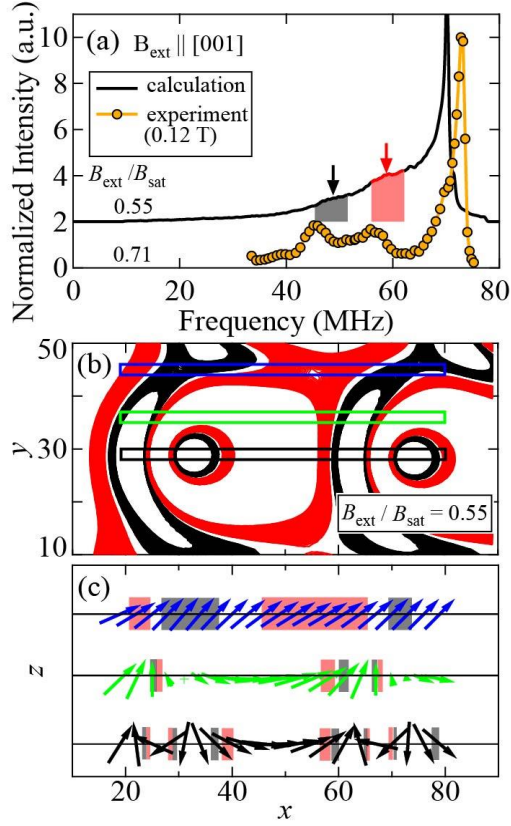

Fig. 15 (a) Comparison between the calculated histogram of the NMR frequency in $B_{ext}/B_{sat} = 0.55$ and the observed spectrum at 0.12 T which corresponds to $B_{ext}/B_{sat} = 0.71$. The histogram is plotted with an offset. (b) Distribution of the NMR frequency around the skewed skyrmions in $B_{ext}/B_{sat} = 0.55$. The black and the red regions show the V sites where the NMR frequency corresponds to the signal of the same color in the calculated histogram shown in (a). (c) Cross sections of the spin configurations in the three regions surrounded by blue, green and black lines shown in (b). The red and black hatches indicate the area in which the V sites contribute to the signal of the same color in the calculated histogram.

## V. Conclusion

We have investigated the magnetic states in the polar magnet $GaV_4Se_8$ by using $^{51}V$ NMR measurements. We have successfully separated the $^{51}V$ NMR signals from the different V sites in the different domains. On the other hand, we find that only the high-frequency region of the NMR spectrum is observed due to a significant increase of the spin-echo decay rate in the low-frequency region.

In $\mathbf{B}_{ext} \parallel [111]$, the structural domains in $GaV_4Se_8$ are divided into the domains in which the polar axis is parallel to or tilted from the magnetic field. In the tilted domains, the NMR signals from the conical state are clearly observed. In sharp contrast, in the parallel domain, we find only the NMR signal from the supermagnetized cycloidal state, but not at all from the skyrmion lattice state. Based on this absence of the NMR signal, we suggest that the spin-echo decay rate is too large to detect the NMR signal in the skyrmion lattice by additional low-energy excitations caused by dynamics of the skyrmions.

In $\mathbf{B}_{\text{ext}} \parallel [001]$, all the structural domains are magnetically equivalent because the angle between the polar axis and the magnetic field is the same for all the domains. Under this oblique field, either the conical state or the skewed skyrmion lattice state is suggested to appear. We compare the observed field dependence of the NMR spectrum with our calculations done for the two states. We cannot distinguish between the two states, because our measurements were limited to high-frequency region due to the enhanced decay rate at lower frequencies. However, the absence of the NMR signal from the skyrmion lattice state in $\mathbf{B}_{\text{ext}} \parallel [111]$ leads us to conclude that the conical state is more likely realized.

**Acknowledgements**
This study was supported by the JSPS KAKENHI Grants No. JP17H02918, No. JP25287083, and No. JP18H04310 (J-Physics).

**Appendix A : Calibration of pulse-conditions for the $^{51}$V NMR measurements**

In this section, we present how to calibrate pulse-conditions for the $^{51}$V NMR measurements in the magnetically ordered states. Let us first show the measurement setups. We used sample #2 which has a tetrahedral crystal shape with an edge size of 1 mm (see Fig. A1). Sample #2 was placed in the NMR coil so that we can apply an oscillation rf field $H_1$ parallel to the $[1\bar{1}0]$ direction as shown in Fig. A1. For detecting spin-echo signals with short spin-echo decay times $T_2$, we used a pulse sequence composed of two identical pulses with a delay time $\tau$ separating these two pulses as short as possible. We set the duration of the first and second rf pulses at 0.8 μs, which suppresses a ringing noise after the second pulse and enables us to reduce the time $\tau$ to 4 μs.

The pulse-condition was calibrated to maximize the spin-echo intensity. Let us focus on the pulse condition in zero external magnetic field at 4.2 K (see Fig. 2). To examine the frequency dependence of the pulse-condition, we checked the pulse-voltage dependence of the spin-echo intensity at different frequencies as shown in Fig. A2. Here, note that the echo intensity is plotted against output power of the first pulse from the NMR power amplifier instead of the pulse-voltage. For the absorption peak observed at 73.8 MHz, the echo intensity is maximized with a pulse-power of 0.08 μJ above which the echo intensity is suppressed with showing the Rabi-oscillation. The optimized pulse-power does not change with lowering resonance frequency down to 60 MHz below which the spin-echo signal is not discernable due to fast spin-echo decay times $T_2$. The pulse-conditions in external magnetic fields were also calibrated in the same way. The optimized pulse-power is gradually increased with the magnetic field, whereas it is hardly changed with frequency in a fixed field.

Remarkably, the optimized pulse-voltage in the ordered state is much smaller than that in the paramagnetic state. For instance, the optimized pulse-voltage in zero external field at 4.2 K is about 130 times smaller than that in the paramagnetic state. This significant pulse-voltage reduction is caused by the ordered magnetic moments as reported in ferromagnets [36]. As discussed for ferromagnets with a weak spin anisotropy, the nuclear spins are excited not only by the external rf field $H_1$ but also the hyperfine field following the electronic moment response to $H_1$. The enhancement factor of $H_1$ is given by the ratio of the hyperfine field to the restoring one, which characterizes the torque experienced by the electronic moment when it is tilted from its position at rest by $H_1$. In the case of the cycloidal state of GaV$_4$Se$_8$, the restoring field is determined by competition between a ferromagnetic exchange and DM interactions, which is overwhelmed by an external magnetic field less than 0.1 T. Taking into account the hyperfine field of ~7 T at the $^{51}$V nucleus, the enhancement factor can be evaluated to be larger than 70, which is compatible with the enhancement factor of 130.

Finally, let us comment about the frequency independence of the pulse-conditions. In noncollinear magnetic states of GaV$_4$Se$_8$, nuclear spins are expected to experience different strength of effective rf field $H_1^{\text{eff}} = H_1 \sin\psi$, depending on the angle $\psi$ between the external rf field and the local magnetic field generated by the ordered magnetic moment. Since the broad NMR spectra observed in magnetically ordered states are caused by distributed directions of the magnetic moments at the V sites, the pulse-condition could depend on the resonance frequency. However, the pulse-condition is hardly changed with the resonance frequency as mentioned above. This is because the region where the angle between the direction of the internal magnetic field and that of $H_1$ changes is limited to a small portion of the sample due to the presence of multiple structural and/or $q$-domains [19,25]. Although a part of NMR signals from the small portion is not detected, it does not affect the overall spectral shape, which allows us to discuss the magnetic states from the spectral shape.

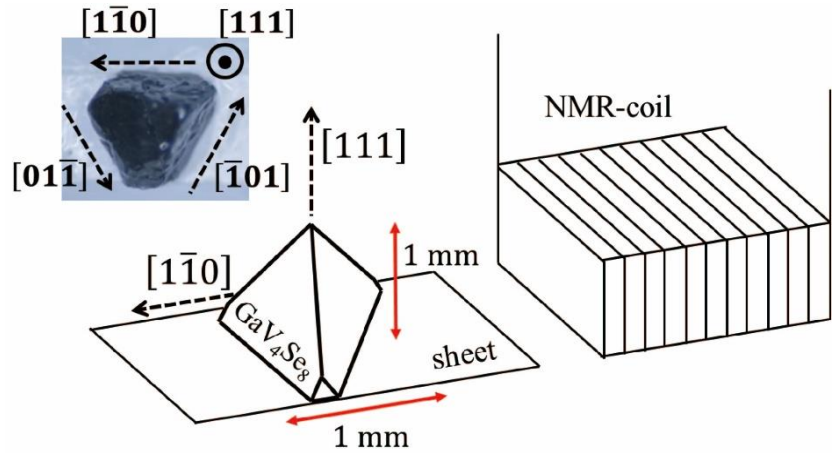

Fig. A1 Schematic illustration of the NMR experimental setting. Sample #2 on a paper sheet is placed in the NMR-coil. The inset shows a photo of Sample #2 viewed along [111] axis.

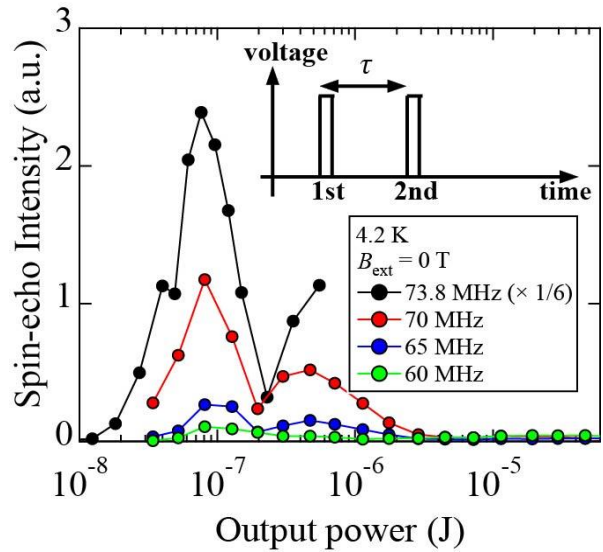

Fig. A2 Pulse-power dependence of the spin-echo intensity at resonance frequencies of 60, 65, 70 and 73.8 MHz measured at 4.2 K in zero external magnetic field. The inset shows the schematic figure of time profile of an NMR pulse sequence.

**Appendix B : $^{51}$V NMR spectra in the paramagnetic phase**

In this section, we show the $^{51}$V NMR spectrum measured in the paramagnetic phase. Figures B1(a) and (b) show the $^{51}$V NMR spectra measured at 30 K in external magnetic fields $B_{\text{ext}}$ applied in the $(1\bar{1}0)$ and $(\bar{1}\bar{1}2)$ planes respectively. Note that the spectra were taken by recording the integrated intensity of the spin-echo signal at a fixed frequency $f_0 = 75.96$ MHz while sweeping the magnetic field. Multiple peaks for the $^{51}$V NMR as well as a set of $^{69}$Ga NMR spectra are observed. In $\mathbf{B}_{\text{ext}} \parallel (1\bar{1}0)$, five sets of $^{51}$V NMR spectra are observed as indicated by arrows. The resonance fields of these spectra shift as the applied field orientation is changed. In $\mathbf{B}_{\text{ext}} \parallel (\bar{1}\bar{1}2)$ five sets of spectra are observed, which show different angle dependence of the resonance fields from those in $\mathbf{B}_{\text{ext}} \parallel (1\bar{1}0)$. In Figs. B1(c) and B1(d), angle dependence of the NMR shifts for these five peaks in $\mathbf{B}_{\text{ext}} \parallel (1\bar{1}0)$ and $\mathbf{B}_{\text{ext}} \parallel (\bar{1}\bar{1}2)$ are shown respectively. The NMR shifts were evaluated by using the equation, $K = f_0/\gamma B_{\text{res}} - 1$. Among the five sets of the NMR spectra, two (one) of them, which are discernable around $\langle 111 \rangle$ axes for $\mathbf{B}_{\text{ext}} \parallel (1\bar{1}0)$ ($\mathbf{B}_{\text{ext}} \parallel (\bar{1}\bar{1}2)$), show quadratic angle dependence (see orange and grey arrows in Figs. B1(a) and B1(b)), while the others show sine curves as a function of the angle. These subpeaks are ascribed to a slight tilting of the direction of the applied magnetic field to the crystal axis or presence of the V sites affected by imperfection of the crystal.

The multiple peaks in the $^{51}$V NMR spectra for $\mathbf{B}_{\text{ext}} \parallel (1\bar{1}0)$ and $\mathbf{B}_{\text{ext}} \parallel (\bar{1}\bar{1}2)$ can be explained by the presence of the four structural domains in which V$_4$ tetrahedra are elongated along four directions of the $\langle 111 \rangle$ axes as shown in Fig. B1(g). Let us assume that a V site for each domain has a uniaxial NMR shift tensor around the elongation (polar) axis. The angle dependence of the NMR shift for the V site in each domain follows the equation,

$$K(\theta_j) = K_{\text{iso}} + K_{\text{ax}}(3\cos^2\theta_j - 1), \tag{B1}$$

where $\theta_j$ is the angle between the polar axis $Z_j$ of the domain $j$ (= A, B, C, and D) and the magnetic-field direction, and $K_{\text{iso}}$ and $K_{\text{ax}}$ are the isotropic and axial anisotropic components of the NMR shift expressed as $K_{\text{iso}} = (2K_\perp + K_\parallel)/3$ and $K_{\text{ax}} = (K_\perp - K_\parallel)/2$, respectively, which results in three (four) sets of absorption peaks for $\mathbf{B}_{\text{ext}} \parallel (1\bar{1}0)$ ($\mathbf{B}_{\text{ext}} \parallel (\bar{1}\bar{1}2)$). All the observed sinusoidal angle dependence of the NMR shifts are well fit to Eq. (B1) with the fitting parameters of $K_{\text{iso}} = -9.76$ % and $K_{\text{ax}} = 4.58$ % as shown by solid lines in Figs. B1(c) and B1(d), which indicates that the observed peaks are assigned to the V sites in the four different structural domains. Note that the experimental data slightly deviate from the fitting curve, which is probably because the NMR shift tensor is not exactly axial symmetric. As shown later, the observed peaks are assigned to V2 sites which have a mirror symmetry $.m$.

To evaluate the hyperfine coupling constants, we measured the NMR spectra in magnetic fields applied along the [111] and [110] directions at various temperatures and obtained temperature dependence of the NMR shifts, $K_\parallel$ and $K_\perp$ as shown in Fig. B2(a). Subsequently, the NMR shift versus magnetic susceptibility plot with temperature as an implicit parameter is obtained as shown in Fig. B2(b). The hyperfine coupling constants are obtained by fitting the data to a linear function $K_{\parallel(\perp)} = A_{\parallel(\perp)}\chi/N_A\mu_B$, where $A_{\parallel(\perp)}$ is the coupling constant, and $N_A$ is the Avogadro constant. The values of the coupling constants are $A_\parallel = -0.13$ T/$\mu_B$ and $A_\perp = -3.47$ T/$\mu_B$.

The NMR spectra observed in the paramagnetic phase are assigned to the V2 sites. To demonstrate this, let us focus on the NMR spectra observed in the magnetically ordered phase at zero external magnetic field. As shown in the Experimental Results (Sec. III C), two sets of NMR spectra assigned to the V1 and V2 sites are observed at 73.7 and 30.7 MHz respectively (see Figs. 2(a) and 2(b)). Both of the observed peaks correspond to the high-frequency edges of double-horn spectra for the cycloidal magnetic state (see Sec. IV A). Since the ordered moments are uniformly distributed in a plane parallel to the principal axis $Z$, the high-frequency edges are given by $f_{\text{edge}(i)} = \gamma|A_{i\perp}\mu|$. Considering that the size of the ordered moment at a V$_4$ tetrahedron is $\mu \sim 0.8\mu_B$, the absolute values of $A_{i\perp}$ estimated from the resonance frequencies in zero external field are $|A_{1\perp}|\sim 8.23$ T/$\mu_B$ and $|A_{2\perp}|\sim 3.43$ T/$\mu_B$, which indicates that the spectra observed in the paramagnetic phase are assigned to the V2 sites. An NMR spectrum from the V1 site is not detected probably due to a short spin-echo decay time.

Even though the V2 site does not have an axial site symmetry, the hyperfine coupling and NMR shift tensors for the V2 site are nearly uniaxial, which must be ascribed to the local charge distribution at the V2 site. As shown in Fig. 1(c), the $a_1$ molecular orbital which an unpaired electron spin occupies is composed of axial charge distributions around the four V sites [27,28]. Such a charge distribution would produce the nearly uniaxial hyperfine field at the $^{51}$V nucleus of the V2 site. Furthermore, from the experimental result that the hyperfine coupling constants for the V1 and V2 sites are different from each other, it is inferred that the charge distribution in the V$_4$ tetrahedron is not uniform. Although the quantitative estimation of the charge distribution is beyond our scope, the larger value of the hyperfine coupling constant for the V1 site suggests that the unpaired electron is more densely distributed at the V1 site in the V$_4$ tetrahedron.

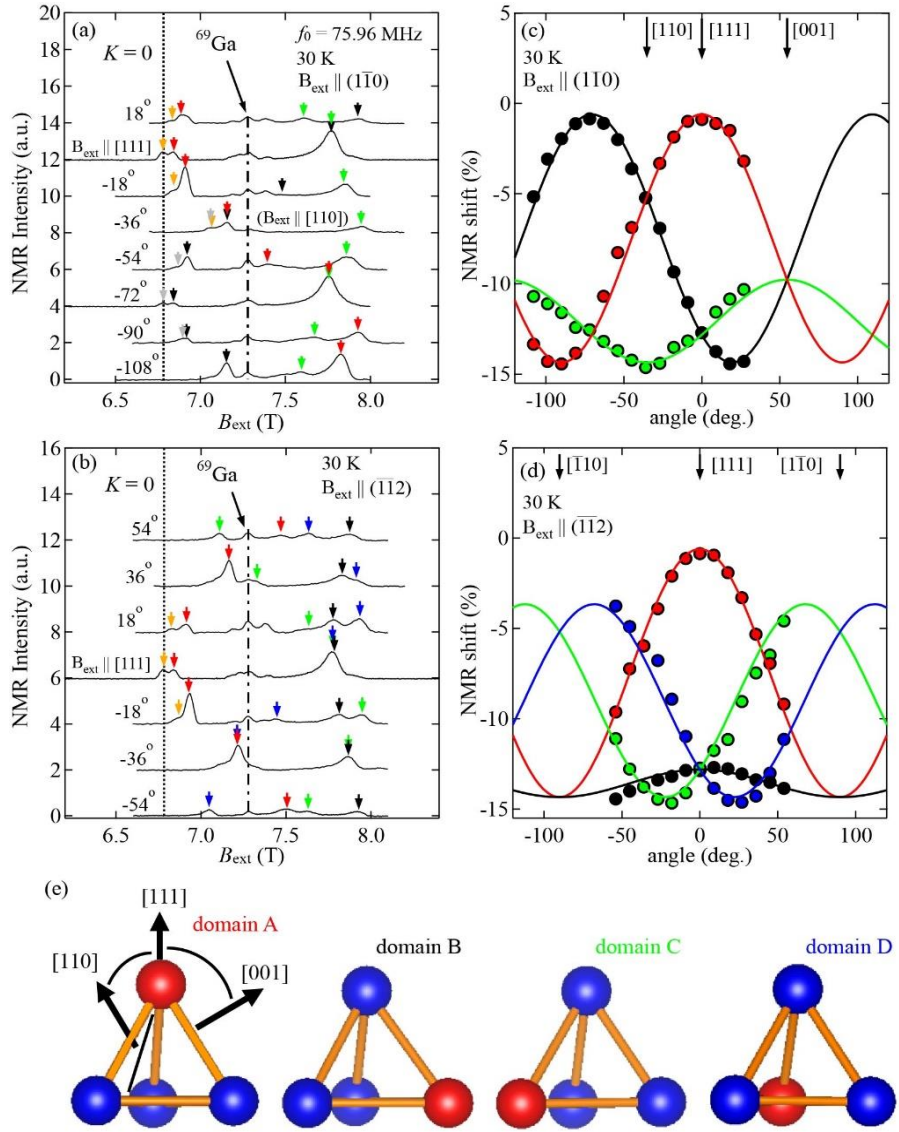

Fig. B1 Magnetic field-swept $^{51}$V NMR spectra measured at 30 K in external magnetic fields applied in the (a) $(1\bar{1}0)$ and (b) $(\bar{1}\bar{1}2)$ planes. The spectra are plotted with offsets. The arrows indicate the peaks of $^{51}$V NMR spectra. The dashed lines represent the magnetic fields corresponding to the NMR shift $K = 0$, while the dash-dotted lines are guides for $^{69}$Ga NMR spectra. Angle dependence of the NMR shifts in the magnetic fields applied in the (c) $(1\bar{1}0)$ and (d) $(\bar{1}\bar{1}2)$ planes. The solid lines are fitting results by $K(\theta) = K_{\text{iso}} + K_{\text{ax}}(3\cos^2\theta - 1)$, where $\theta$ indices the angle between the principal axis for each domain and magnetic field. (e) Schematic illustrations of the V$_4$ tetrahedra in the four structural domains A-D drawn by VESTA [26].

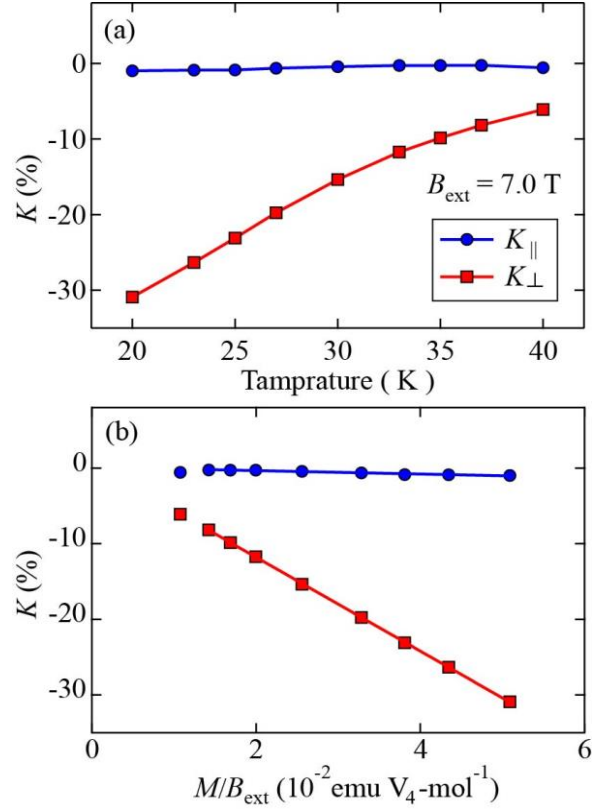

Fig. B2 (a) Temperature dependence of the NMR shifts $K_\parallel$ and $K_\perp$ measured in external field of 7 T applied along the [111] and [110] axes. (b) NMR shift versus magnetic susceptibility plots. The solid lines are fitting results by using a linear function.

**Appendix C : Evaluation of the hyperfine coupling constants for the V1 site**

In this section we aim to evaluate the hyperfine coupling constants for the V1 site from the NMR signal in the forced-ferromagnetic state for $\mathbf{B}_{\text{ext}} \parallel [110]$ and $\mathbf{B}_{\text{ext}} \parallel [001]$. First, let us focus on the spectra in the external magnetic field $\mathbf{B}_{\text{ext}} \parallel [110]$. In the magnetic field $\mathbf{B}_{\text{ext}} \parallel [110]$, we observed two $^{51}$V NMR spectra assigned to the V1 site at the structural domains B, D and A, C as shown in Fig. C1(a). The spectrum for the domains B, D shifts to low frequency linearly with increasing magnetic field. Since the applied field is perpendicular to the polar axis, the resonance frequency of the spectral peak follows the equation

$$f_{\text{res}} = \gamma |B_{\text{ext}} + A_{1\perp}\mu|. \tag{C1}$$

Our experimental data $f_{\text{res}}(B_{\text{ext}})$ can be fit to this equation with $A_{1\perp}\mu = -6.7$ T as shown by the black solid line in Fig. C1(b). On the other hand, the spectrum for the domains A, C once shifts to low frequency up to 3 T, followed by shifting to high frequency above this field as shown in Figs. B1(a) and B1(b). To explain this magnetic-field dependence of the resonance frequency let us focus on domain A. The resonance frequency of the spectral peak at the high field region where the ordered magnetic moments are parallel to the direction of the magnetic field should follow the equation

$$f_{res} = \gamma \left| B_{ext} \begin{pmatrix} -1/\sqrt{3} \\ 0 \\ \sqrt{2}/\sqrt{3} \end{pmatrix} + \begin{pmatrix} A_{1\perp} & & \\ & A_{1\perp} & \\ & & A_{1\parallel} \end{pmatrix} \cdot \mu \begin{pmatrix} -1/\sqrt{3} \\ 0 \\ \sqrt{2}/\sqrt{3} \end{pmatrix} \right|. \quad (C2)$$

Above 4 T, the observed resonance frequency can be fit to this equation with $A_{1\perp}\mu = -6.7$ T and $A_{1\parallel}\mu = -1.1$ T. The deviation between the observed frequency and the equation below 3 T is due to the tilting of the ordered magnetic moment from the direction of the magnetic-field. The tilting angle, which is defined as an angle between the ordered moment and the normal to the cycloidal plane (see Figs. 9(d) and 9(e)), approaches the saturation value with increasing magnetic field as shown in Fig. C1(d). Given that the size of the ordered magnetic moment $\mu$ per single V4 tetrahedron is $0.8\,\mu_B$, the hyperfine coupling constants $A_{1\perp}$ and $A_{1\parallel}$ are estimated as $-8.38\,\text{T}/\mu_B$ and $-1.38\,\text{T}/\mu_B$.

By using the coupling constants obtained above, magnetic-field dependence of the resonance frequencies $f_{res}(B_{ext})$ in $\mathbf{B}_{ext} \parallel [001]$ can be explained as follows. As mentioned in the main text, we observed a single peak of the $^{51}$V NMR spectrum assigned to the V1 site in $\mathbf{B}_{ext} \parallel [001]$, because the direction of the applied magnetic field is equivalent to the four structural domains, i.e. the applied field is tilted by $54.7°$ from the polar axis of each structural domain. The resonance frequency $f_{res}(B_{ext})$ is presented in Fig. C1(c). At high magnetic field above ~2 T, the frequency $f_{res}(B_{ext})$ follows an equation $f_{res} = \gamma|\mathbf{B}_{ext} + \mathbf{A}_1\boldsymbol{\mu}|$ with the coupling constants $A_{1\perp} = -8.38\,\text{T}/\mu_B$, and $A_{1\parallel} = -1.38\,\text{T}/\mu_B$ as shown by the solid line in Fig. C1(c). The deviation between the observed and calculated frequencies below 2 T is due to a tilting angle between the direction of magnetic-field and ordered moment (see Fig. C1(d)), which is caused by an easy-plane-type anisotropy of the magnetic moment.

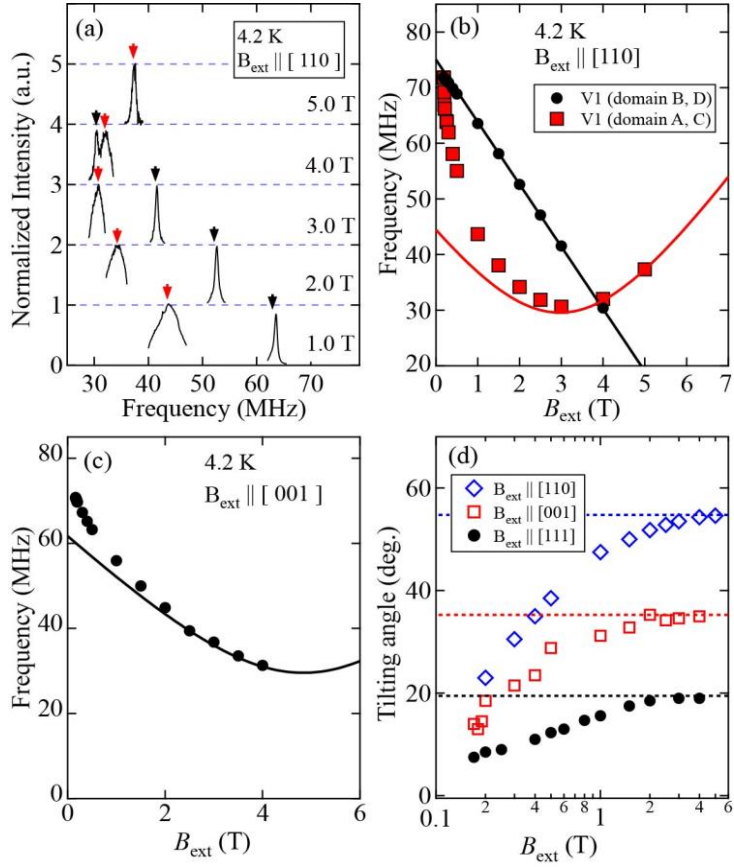

Fig. C1 (a) $^{51}$V NMR spectra assigned to the V1 sites in the domains A,C and B, D measured in the external magnetic field $\mathbf{B}_{\text{ext}} \parallel [110]$ at 4.2 K. The spectra are plotted with offsets. Magnetic-field dependence of the resonance frequencies for the $^{51}$V NMR spectral peaks in (b) $\mathbf{B}_{\text{ext}} \parallel [110]$ and (c) $\mathbf{B}_{\text{ext}} \parallel [001]$. The solid lines are calculated resonance frequencies with an assumption that the ordered magnetic moments are parallel to the magnetic field and the coupling constants follow the relations $A_{1\perp} = -8.38 \text{ T}/\mu_B$ and $A_{1\parallel} = -1.38 \text{ T}/\mu_B$. (d) Magnetic-field dependence of the tilting angles between the directions of the magnetic moment and magnetic field in the fields applied along three axes [111], [110] and [001]. The dotted lines in (d) show the tilting angles when the magnetic moment becomes parallel to the applied field.

## Appendix D : Calculated $^{51}$V NMR spectrum for the SkL state

The Néel-type skyrmion lattice with incommensurate propagation vectors produces a peculiar histogram of the $^{51}$V NMR frequencies as shown in Fig. 8(b). In this magnetic state, a unit vector for the orientation of each magnetic moment at a position of $\mathbf{r}_i$ is expressed as

$$\mathbf{n}(\mathbf{r}_i) = N_i \left[ \sum_{j=1}^{3} \left\{ \mathbf{e}_z \cos(\mathbf{q}_j \cdot \mathbf{r}_i) + \mathbf{e}_{q_j} \sin(\mathbf{q}_j \cdot \mathbf{r}_i) \right\} \right], \tag{D1}$$

where $N_i$ is a parameter for normalizing the vector, $\mathbf{q}_j$ are propagation vectors, $\mathbf{q}_1 = q(1,0,0)$, $\mathbf{q}_2 = q(-1/2, \sqrt{3}/2, 0)$ and $\mathbf{q}_3 = q(-1/2, -\sqrt{3}/2, 0)$, and $\mathbf{e}_z$ and $\mathbf{e}_{q_j}$ are unit vectors following $\mathbf{e}_z = (0,0,1)$ and $\mathbf{e}_{q_j} = \mathbf{q}_j/q_j$. The $^{51}$V NMR frequency at the V1 site is given by $f_{\text{res}}(\mathbf{r}_i) = \gamma |\mathbf{A}_1 \cdot \mu \mathbf{n}(\mathbf{r}_i)|$. As the histogram of these frequencies we obtain the spectrum shown in Fig. 8(b).

**References**

[1] I. Dzyaloshinsky, A thermodynamic theory of "weak" ferromagnetism of antiferromagnetics, Journal of Physics and Chemistry of Solids **4**, 241 (1958).

[2] T. Moriya, Anisotropic Superexchange Interaction and Weak Ferromagnetism, Phys. Rev. **120**, 91 (1960).

[3] A. Bogdanov and A. Hubert, Thermodynamically stable magnetic vortex states in magnetic crystals, Journal of Magnetism and Magnetic Materials **138**, 255 (1994).

[4] Y. Tokura and N. Kanazawa, Magnetic Skyrmion Materials, Chem. Rev. **121**, 2857 (2021).

[5] A. Neubauer, C. Pfleiderer, B. Binz, A. Rosch, R. Ritz, P. G. Niklowitz, and P. Böni, Topological Hall Effect in the A Phase of MnSi, Phys. Rev. Lett. **102**, 186602 (2009).

[6] M. Lee, W. Kang, Y. Onose, Y. Tokura, and N. P. Ong, Unusual Hall Effect Anomaly in MnSi under Pressure, Phys. Rev. Lett. **102**, 186601 (2009).

[7] T. Kurumaji, T. Nakajima, M. Hirschberger, A. Kikkawa, Y. Yamasaki, H. Sagayama, H. Nakao, Y. Taguchi, T. Arima, and Y. Tokura, Skyrmion lattice with a giant topological Hall effect in a frustrated triangular-lattice magnet, Science **365**, 914 (2019).

[8] A. Fert, V. Cros, and J. Sampaio, Skyrmions on the track, Nature Nanotech **8**, 152 (2013).

[9] N. S. Kiselev, A. N. Bogdanov, R. Schäfer, and U. K. Rößler, Chiral skyrmions in thin magnetic films: new objects for magnetic storage technologies?, J. Phys. D: Appl. Phys. **44**, 392001 (2011).

[10] N. Nagaosa and Y. Tokura, Topological properties and dynamics of magnetic skyrmions, Nature Nanotech **8**, 899 (2013).

[11] Y. Tokura, S. Seki, and N. Nagaosa, Multiferroics of spin origin, Rep. Prog. Phys. **77**, 076501 (2014).

[12] S. Mühlbauer, B. Binz, F. Jonietz, C. Pfleiderer, A. Rosch, A. Neubauer, R. Georgii, and P. Böni, Skyrmion Lattice in a Chiral Magnet, Science **323**, 915 (2009).

[13] M. Uchida, N. Nagaosa, J. P. He, Y. Kaneko, S. Iguchi, Y. Matsui, and Y. Tokura, Topological spin textures in the helimagnet FeGe, Phys. Rev. B **77**, 184402 (2008).

[14] X. Z. Yu, Y. Onose, N. Kanazawa, J. H. Park, J. H. Han, Y. Matsui, N. Nagaosa, and Y. Tokura, Real-space observation of a two-dimensional skyrmion crystal, Nature **465**, 901 (2010).

[15] S. Seki, X. Z. Yu, S. Ishiwata, and Y. Tokura, Observation of Skyrmions in a Multiferroic Material, Science **336**, 198 (2012).

[16] I. Kézsmárki et al., Néel-type skyrmion lattice with confined orientation in the polar magnetic semiconductor $GaV_4S_8$, Nature Mater **14**, 1116 (2015).

[17] E. Ruff, S. Widmann, P. Lunkenheimer, V. Tsurkan, S. Bordács, I. Kézsmárki, and A. Loidl, Multiferroicity and skyrmions carrying electric polarization in $GaV_4S_8$, Science Advances **1**, e1500916 (2015).

[18] J. S. White, Á. Butykai, R. Cubitt, D. Honecker, C. D. Dewhurst, L. F. Kiss, V. Tsurkan, and S. Bordács, Direct evidence for cycloidal modulations in the thermal-fluctuation-stabilized spin spiral and skyrmion states of $GaV_4S_8$, Phys. Rev. B **97**, 020401 (2018).

[19] S. Bordács et al., Equilibrium Skyrmion Lattice Ground State in a Polar Easy-plane Magnet, Sci Rep **7**, 7584 (2017).

[20] Y. Fujima, N. Abe, Y. Tokunaga, and T. Arima, Thermodynamically stable skyrmion lattice at low temperatures in a bulk crystal of lacunar spinel $GaV_4Se_8$, Phys. Rev. B **95**, 180410 (2017).

[21] M. Akazawa, H.-Y. Lee, H. Takeda, Y. Fujima, Y. Tokunaga, T. Arima, J. H. Han, and M. Yamashita, Topological thermal Hall effect of magnons in magnetic skyrmion lattice, Phys. Rev. Research **4**, 043085 (2022).

[22] T. Kurumaji, T. Nakajima, V. Ukleev, A. Feoktystov, T. Arima, K. Kakurai, and Y. Tokura, Néel-Type Skyrmion Lattice in the Tetragonal Polar Magnet $VOSe_2O_5$, Phys. Rev. Lett. **119**, 237201 (2017).

[23] K. Geirhos et al., Macroscopic manifestation of domain-wall magnetism and magnetoelectric effect in a Néel-type skyrmion host, Npj Quantum Mater. **5**, 1 (2020).

[24] B. Gross, S. Philipp, K. Geirhos, A. Mehlin, S. Bordács, V. Tsurkan, A. Leonov, I. Kézsmárki, and M. Poggio, Stability of Néel-type skyrmion lattice against oblique magnetic fields in $GaV_4S_8$ and $GaV_4Se_8$, Phys. Rev. B **102**, 104407 (2020).

[25] M. Prinz-Zwick, T. Gimpel, K. Geirhos, S. Ghara, C. Steinbrecht, V. Tsurkan, N. Büttgen, and I. Kézsmárki, Probing multiferroic order parameters and domain population

via nuclear spins, Phys. Rev. B **105**, 014301 (2022).

[26] K. Momma and F. Izumi, *VESTA 3* for three-dimensional visualization of crystal, volumetric and morphology data, J Appl Crystallogr **44**, 1272 (2011).

[27] R. Pocha, D. Johrendt, and R. Pöttgen, Electronic and Structural Instabilities in $GaV_4S_8$ and $GaMo_4S_8$, Chem. Mater. **12**, 2882 (2000).

[28] H. Nakamura, H. Chudo, and M. Shiga, Structural transition of the tetrahedral metal cluster: nuclear magnetic resonance study of $GaV_4S_8$, J. Phys.: Condens. Matter **17**, 6015 (2005).

[29] P. Mischo, F. Decker, U. Häcker, K.-P. Holzer, J. Petersson, and D. Michel, Low-Frequency Phason and Amplitudon Dynamics in the Incommensurate Phase of $Rb_2ZnCl_4$, Phys. Rev. Lett. **78**, 2152 (1997).

[30] B. Topi, U. Haeberlen, and R. Blinc, $^{39}$K NMR evidence for a phason gap in $K_2SeO_4$, Phys. Rev. B **40**, 799 (1989).

[31] N. Higa et al., Critical slowing-down and field-dependent paramagnetic fluctuations in the skyrmion host EuPtSi: $\mu$SR and NMR studies, Phys. Rev. B **104**, 045145 (2021).

[32] D. Ehlers, I. Stasinopoulos, V. Tsurkan, H.-A. Krug Von Nidda, T. Fehér, A. Leonov, I. Kézsmárki, D. Grundler, and A. Loidl, Skyrmion dynamics under uniaxial anisotropy, Phys. Rev. B **94**, 014406 (2016).

[33] A. Vansteenkiste, J. Leliaert, M. Dvornik, M. Helsen, F. Garcia-Sanchez, and B. Van Waeyenberge, The design and verification of MuMax3, AIP Advances **4**, 107133 (2014).

[34] A. Štefančič, S. J. R. Holt, M. R. Lees, C. Ritter, M. J. Gutmann, T. Lancaster, and G. Balakrishnan, Establishing magneto-structural relationships in the solid solutions of the skyrmion hosting family of materials: $GaV_4S_{8-y}Se_y$, Sci Rep **10**, 9813 (2020).

[35] P. Padmanabhan, F. Sekiguchi, R. B. Versteeg, E. Slivina, V. Tsurkan, S. Bordács, I. Kézsmárki, and P. H. M. van Loosdrecht, Optically Driven Collective Spin Excitations and Magnetization Dynamics in the Néel-type Skyrmion Host $GaV_4S_8$, Phys. Rev. Lett. **122**, 107203 (2019).

[36] C. Meny and P. Panissod, *Nuclear Magnetic Resonance in Ferromagnets: Ferromagnetic Nuclear Resonance; a Very Broadband Approach*, in *Annual Reports on NMR Spectroscopy*, Vol. 103 (Elsevier, 2021), pp. 47–96.